# HDL and plaque regression in a multiphase model of early atherosclerosis

Ishraq U. Ahmed[1] and Mary R. Myerscough[1]

[1]School of Mathematics and Statistics, University of Sydney

December 19, 2023

### Abstract

Atherosclerotic plaques are accumulations of cholesterol-engorged macrophages in the artery wall. Plaque growth is initiated and sustained by the deposition of low density lipoproteins (LDL) in the artery wall. High density lipoproteins (HDL) counterbalance the effects of LDL by accepting cholesterol from macrophages and removing it from the plaque. In this paper, we develop a free boundary multiphase model to investigate the effects of LDL and HDL on early plaque development. We examine how the rates of LDL and HDL deposition affect cholesterol accumulation in macrophages, and how this impacts cell death rates and emigration. We identify a region of LDL-HDL parameter space where plaque growth stabilises for low LDL and high HDL influxes, due to macrophage emigration and HDL clearance that counterbalances the influx of new cells and cholesterol. We explore how the efferocytic uptake of dead cells and the recruitment of new macrophages affect plaque development for a range of LDL and HDL levels. Finally, we consider how changes in the LDL-HDL profile can change the course of plaque development. We show that changes towards lower LDL and higher HDL can slow plaque growth and even induce regression. We find that these changes have less effect on larger, more established plaques, and that temporary changes will only slow plaque growth in the short term.

## 1 Introduction

Atherosclerosis is a chronic inflammatory disease where lipids and lipid-engorged cells accumulate in the wall of major arteries. This accumulation of material can cause arterial narrowing and other complications. Atherosclerosis is a key cause of cardiovascular disease-related deaths (3; 4). Low density lipoproteins (LDL) and their byproducts play a central role in atherosclerosis. The deposition of LDL in the artery wall and their subsequent uptake by macrophages drive the early inflammatory response (37). High density lipoproteins (HDL) in contrast are known to be atheroprotective, as they accept excess cholesterol from macrophages and upregulate anti-inflammatory behaviour (4).

In this paper, we develop a multiphase model for early atherosclerotic plaque growth that focuses on macrophage-lipoprotein interactions, and tracks levels of intracellular cholesterol. Our model includes spatial crowding effects, and a variable plaque size. We use this model to investigate the competing effects of LDL and HDL, and how these interaction affect the spatial structure and growth of early plaques.

Atherosclerotic lesions start to form when cholesterol-carrying low density lipoproteins (LDL) from the bloodstream deposit in the intima (37; 20). The intima is a thin layer that separates the layer of cells that line the inside of the artery (the endothelium) from the muscular outer layer of the artery wall (the media). LDL particles in the intima undergo chemical modification, principally via oxidation, to produce modified LDL (modLDL) (39; 67). Intimal modLDL stimulates an immune response which draws monocytes from the bloodstream into the artery wall, where they differentiate into macrophages (7; 42). Macrophages take up modLDL via phagocytosis, and the resulting cholesterol-engorged cells are referred to as macrophage foam cells (68). Lipid and cellular material can continue to accumulate over years, and plaques may become unstable and prone to rupture. When rupture occurs, the material that escapes from the plaque can trigger the formation of a blood clot or thrombus, which, in the worst case, can cause a myocardial infarction or stroke (37).

Macrophages and other cells constantly undergo a normal form of programmed cell death called apoptosis. Apoptotic cells express find-me and eat-me signals to promote their clearance by live macrophages, which consume dying cells via efferocytosis (56). In healthy efferocytosis, engulfing macrophages produce anti-inflammatory cytokines to prevent runaway inflammation. Macrophage foam cells can also emigrate from the intima via lymphatic vessels that connect to the outer artery layers, and thereby carry lipid out of the plaque (35; 5).

In atherosclerotic plaques, macrophage anti-inflammatory behaviours can become impaired due to excessive cholesterol loading from modLDL (57; 68). Macrophage foam cells often accumulate toxic levels of intracellular cholesterol, which triggers apoptotic pathways and impairs their efferocytic ability (54). Macrophages exposed to modLDL produce fewer anti-inflammatory cytokines and higher quantities of pro-inflammatory cytokines that attract new monocytes (69). Foam cells also exhibit decreased expression of receptors that are known to be involved with emigration in dendritic cells (50), though the mechanisms of this are less well understood. As a result, inflammation can become chronic, so that plaques accumulate a mass of dead and dying cells that overwhelms the reduced efferocytic capacity of live macrophages. Apoptotic cells that are not cleared quickly undergo an uncontrolled form of cell death called necrosis, where cells break down and release their contents into the plaque space. Necrotic macrophages produce "find-me" and "eat-me" signals at much lower levels than apoptotic cells, and are ingested much less readily. Necrotic material promotes the production of pro-inflammatory cytokines by foam cells and endothelial cells, which attracts more macrophages that are then likely to undergo necrosis. A defining feature of advanced plaques is the presence of a large necrotic core that consists of lipids and debris released by necrotic cells (53).

High density lipoproteins (HDL) are known to possess numerous atheroprotective properties (52; 4; 9). These include HDL's role in reverse cholesterol transport, where HDL particles accept excess cholesterol from macrophages and remove it from the plaque, most likely via the lymphatic vasculature (40). HDL also inhibits the oxidation of LDL, and the production of endothelial cell adhesion molecules and pro-inflammatory cytokines. HDL facilitates macrophage clearance by upregulating the CCR7 receptor, which promotes emigration in monocyte-derived cells (61).

The atherogenic effects of high LDL and low HDL levels have been well established in animal models (62; 32). In experiments by Feig et al (21), plaque-bearing aortic tissue from apoE$^{-/-}$ mice, which are genetically engineered to be susceptible to atherosclerosis, was transplanted into various groups of recipient mice whose blood lipid profiles had lower LDL, higher HDL, or both. Plaque foam cells taken from the high HDL groups exhibited decreased expression of pro-inflammatory markers, and enriched expression of markers associated with cell emigration and tissue repair. Plaque size was also observed to regress post-transplant in the recipient groups, but not in the control apoE$^{-/-}$ group.

In mathematical models of plaque growth, a common feature is the inclusion of LDL as a pro-atherogenic species that attracts monocytes and/or macrophages, and promotes the formation of foam cells (48). Earlier ODE and spatial models of early plaque formation treat new macrophages and cholesterol-laden foam cells as two distinct but phenotypically homogeneous species. Recent non-spatially structured population models include macrophage internalised lipid load as a continuous structural variable (25; 14; 64). In these models, the rates of cell death, emigration, modLDL phagocytosis, and efferocytosis may depend functionally on the intracellular lipid variable.

Some models include the atheroprotective properties of HDL, including ODE models (19; 59), spatially structured PDE models (12), and cellular lipid content-structured PDE models (24; 14). In these models, lower influxes of LDL and higher influxes of HDL into the plaque are associated with lower overall foam cell counts and slower rates of plaque growth, measured by the rate of accumulation of new cells. In models that include macrophage emigration, increasing HDL can enable cell numbers either to reach steady state or to decrease. This can happen due to direct HDL upregulation of emigration (19), or by delaying necrosis due to excess cholesterol accumulation, thereby allowing foam cells to emigrate before they die (14).

Most plaque models with HDL lack spatial structure or model the intima as a fixed spatial domain, and use total cell numbers as a proxy for plaque size. Some models (27; 1) employ a multiphase framework with a moving boundary. This provides a natural framework to represent volume exclusion by plaque constituents, and permits the domain size to change in response to the net influx or outflow of plaque constituents. In these models, plaque size can be directly identified with the size of the spatial domain. In Ahmed et al (1), we presented a simple multiphase model for early plaques that includes macrophages, modLDL, and dead cells. Although this model does not include HDL, or distinguish macrophage behaviour based on cholesterol content, it provides a good foundation for further development.

In this paper, we extend our earlier model (1) to include continuous spatial variation in intracellular cholesterol, and HDL's inhibitory effects on LDL oxidation and monocyte recruitment. This allows us to model continuous variation in foam cell behaviour caused by cholesterol toxicity. The paper is structured as follows. Section 2 describes the full model, including model parameters and numerical solution. In Section 3, we consider plaque growth without HDL, and the influence of rates of efferocytosis, monocyte recruitment, and foam cell emigration on plaque development. In Section 4, we consider a submodel that only includes HDL's cholesterol efflux capabilities. In Section 5, we use the full model to explore the effect of the influx of LDL and HDL on plaque development. We identify regions of LDL-HDL space with stable plaque size and with unbounded plaque growth, as well as a borderline region where small changes in efferocytosis or monocyte recruitment can tip a plaque from the unstable to the stable region or vice versa. In Section 6, we look at how time variations in LDL and HDL influx affect plaque development. We show that sudden changes to a lower LDL influx and/or higher HDL influx can reduce cholesterol accumulation and induce plaque regression. We also show that the effectiveness of these changes depends on when they are administered, and whether they are sustained over time. We conclude in Section 7 with a discussion of the results.

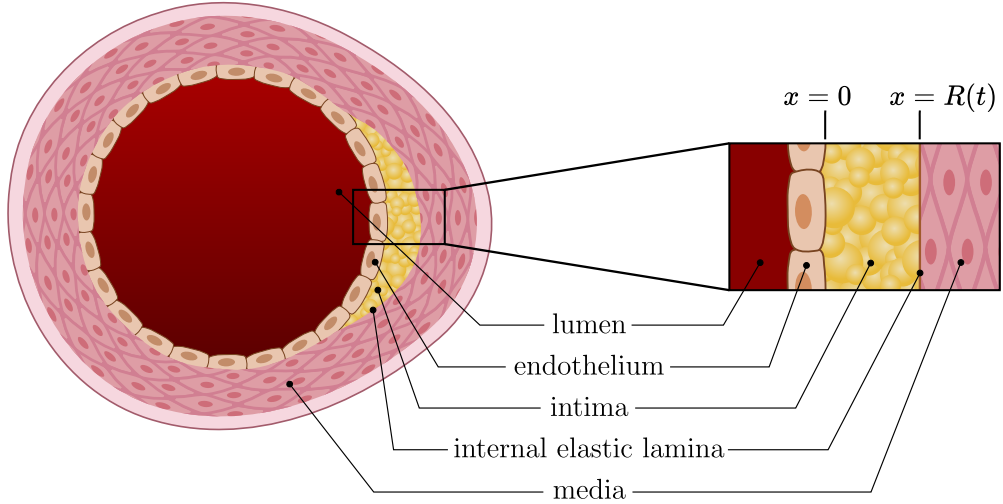

Figure 1: A 2-dimensional cross section of an artery, with a 1-dimensional radial cross section taken through the plaque for the problem domain.

## 2 Model formulation

We represent the intima as a 1-dimensional Cartesian domain, bounded by the endothelium at $x = 0$, and by the elastic lamina (which separates the intima from the media) at $x = R(t)$. The domain here represents a radial cross-section through the artery wall (see Figure 1). We allow the $\tilde{x} = \tilde{R}(\tilde{t})$ boundary to vary with time to allow intimal growth. The $x = 0$ boundary is kept fixed, and we model growth relative to this domain boundary. We also assume that the artery diameter is much larger than the width of the intima. This is true for early stage plaques (8), hence Cartesian coordinates are a suitable approximation for the model. We notate all quantities in the full dimensional model with tildes, which we remove for the nondimensional model in Section 2.1.

Plaque tissue is represented as a multiphase mixture of macrophage foam cells, lipoproteins,

and dead cells, as in Ahmed et al (1). In this model we include both HDL and LDL. We model foam cells as a combination of two distinct phases: macrophages and their intracellular cholesterol content. We use "foam cells" to refer in aggregate to the sum of these two phases. Similarly, we model dead cellular material as two phases: cholesterol from dead cells, and dead material deriving from other cell constituents.

The model contains the macrophage foam cell phases:

- $\tilde{m}(\tilde{x}, \tilde{t})$: macrophages, including the volume of proteins, cytoplasm, etc, as well as endogenous lipid, but excluding their accumulated cholesterol content. This phase encompasses all monocyte-derived cells, including M1 and M2 macrophages and dendritic cells. (43),

- $\tilde{a}(\tilde{x}, \tilde{t})$: accumulated intracellular cholesterol, which is separate from foam cells' other constituents.

The model contains the following lipoprotein phases:

- $\tilde{l}(\tilde{x}, \tilde{t})$: chemically modified LDL, i.e. modLDL,

- $\tilde{h}(\tilde{x}, \tilde{t})$: cholesterol-saturated HDL particles,

- $\tilde{p}(\tilde{x}, \tilde{t})$: nascent (unsaturated) HDL particles that are able to receive cholesterol,

and the dead cell phases:

- $\tilde{b}(\tilde{x}, \tilde{t})$: cholesterol from dead cells,

- $\tilde{c}(\tilde{x}, \tilde{t})$: other dead cellular material (from cell proteins, cytoplasm, etc.).

For modLDL and saturated HDL particles, we only consider the volume of the cholesterol being transported, and ignore the volume of the apolipoprotein and phospholipid outer layer. We therefore treat $\tilde{p}$ as a non-volume-occupying species, and $\tilde{h}$ as a volume-occupying phase. The cellular phases $\tilde{m}$ and $\tilde{c}$ and the remaining cholesterol phases $\tilde{a}$ and $\tilde{b}$ are volume-occupying. The density of each volume-occupying phase is measured in cells or equivalent cell volumes per unit length. We consider $\tilde{p}$ to represent the maximum cholesterol capacity of the unsaturated HDL particles at a given position. For nondimensionalisation purposes, we give $\tilde{p}$ the same units as the volume-occupying phases. We assume there are no voids throughout the plaque, so that the volume-occupying phases obey a no-voids condition

$$\tilde{m} + \tilde{c} + \tilde{l} + \tilde{a} + \tilde{b} + \tilde{h} = N_0 \tag{1}$$

everywhere, where $N_0$ denotes the total phase density, which is constant with time and space.

The species obey continuity equations

$$\frac{\partial \tilde{u}}{\partial \tilde{t}} = -\frac{\partial}{\partial \tilde{x}}(\tilde{J}_u + \tilde{v}\tilde{u}) + \tilde{s}_u, \tag{2}$$

for species $\tilde{u} = \tilde{m}, \tilde{a}, \tilde{l}, \tilde{c}, \tilde{b}, \tilde{h}, \tilde{p}$. The total flux for species $\tilde{u}$ is separated into a bulk advection term $\tilde{v}\tilde{u}$, where $\tilde{v}$ is the bulk mixture velocity, and a phase-specific flux term $\tilde{J}_u$ that describes interphase motion due to random motion or chemotaxis for instance. Each species also has a species-specific mass exchange term $\tilde{s}_u$ that includes source and sink terms.

**Mass exchange terms**

To model the effects of free cholesterol cytotoxicity on macrophage function, we define the foam cell lipid fraction

$$\bar{a} = \frac{\tilde{a}}{\tilde{m} + \tilde{a}}, \tag{3}$$

which defines the amount of intracellular cholesterol as a fraction of the local foam cell volume. We note that $\bar{a}$ is bounded by $0 \leq \bar{a} < 1$ (assuming $\tilde{m} > 0$ everywhere), and approaches 1 as foam cells become engorged on cholesterol.

The two cellular phases, namely macrophages $\tilde{m}$ and dead cells $\tilde{c}$, undergo mass exchange due to macrophage death and efferocytosis. We ignore the distinction between apoptotic and necrotic cell death, and use a generic death term to encompass both processes. We model free cholesterol toxicity by assuming that the macrophage death rate increases linearly with the intracellular lipid fraction. Macrophages ingest dead material via efferocytosis, which we model with a mass-action term. The corresponding mass exchange terms for the cellular phases are

$$\tilde{s}_m = -\tilde{\mu}_a \bar{a}\tilde{m} + \tilde{\mu}_e \tilde{m}\tilde{c}, \tag{4}$$
$$\tilde{s}_c = +\tilde{\mu}_a \bar{a}\tilde{m} - \tilde{\mu}_e \tilde{m}\tilde{c}, \tag{5}$$

where $\tilde{\mu}_a$ is the maximum rate of macrophage death, and $\tilde{\mu}_e$ is the rate of efferocytosis per macrophage per unit of available dead material. We note that the cellular phases obey local mass conservation since $\tilde{s}_m + \tilde{s}_c = 0$.

The cholesterol phases, namely modLDL, HDL, and intracellular cholesterol from live and dead cells, also undergo mass exchange due to macrophage death, efferocytosis, modLDL phagocytosis, and HDL cholesterol efflux. Macrophages ingest modLDL via phagocytosis, which is modelled using a mass-action term. Cholesterol from modLDL is directly internalised into the intracellular cholesterol phase $\tilde{a}$. Macrophages also ingest cholesterol from dead cells $\tilde{b}$ via efferocytosis, which likewise feeds back into $\tilde{a}$. The efferocytic uptake of $\tilde{b}$ occurs at the same rate as $\tilde{c}$. Cell death causes intracellular cholesterol to be released into the corresponding dead material phase at the same rate as cellular material. Foam cells offload cholesterol onto unsaturated HDL particles, which results in their conversion into saturated HDL. We model cholesterol efflux from foam cells using a mass-action term between intracellular cholesterol and the unsaturated HDL capacity. Unsaturated HDL particles will also undergo oxidative degradation and lose their capacity to remove cholesterol from cells. We model this degradation with a linear decay term. The cholesterol phase mass exchange terms are given by

$$\tilde{s}_l = -\tilde{\mu}_p \tilde{m}\tilde{l}, \tag{6}$$
$$\tilde{s}_a = +\tilde{\mu}_p \tilde{m}\tilde{l} + \tilde{\mu}_e \tilde{m}\tilde{b} - \tilde{\mu}_a \bar{a}\tilde{a} - \tilde{\mu}_h \tilde{a}\tilde{p}, \tag{7}$$
$$\tilde{s}_b = -\tilde{\mu}_e \tilde{m}\tilde{b} + \tilde{\mu}_a \bar{a}\tilde{a}, \tag{8}$$
$$\tilde{s}_h = +\tilde{\mu}_h \tilde{a}\tilde{p}, \tag{9}$$
$$\tag{10}$$

and the sole non-volume-occupying species' mass exchange term is

$$\tilde{s}_p = -\tilde{\mu}_h \tilde{a} \tilde{p} - \tilde{\delta}_p \tilde{p}. \tag{11}$$

In the above, $\tilde{\mu}_a$ is the maximum rate of macrophage death, $\tilde{\mu}_p$ and $\tilde{\mu}_e$ are the rates of lipid phagocytosis and efferocytosis per macrophage per unit of available modLDL or dead lipid material respectively, and $\tilde{\mu}_h$ is the rate of cholesterol efflux per unit of available intracellular cholesterol and unsaturated HDL. Like the cellular phases, the cholesterol phases obey local mass conservation since $\tilde{s}_l + \tilde{s}_a + \tilde{s}_b + \tilde{s}_h = 0$.

**Flux terms and boundary conditions**

For convenience, we define the individual phase velocity $\tilde{v}_u$ for a phase $\tilde{u}$ using

$$\tilde{v}_u \tilde{u} = \tilde{J}_u + \tilde{v} \tilde{u}. \tag{12}$$

The continuity equation for phase $\tilde{u}$ can then be rephrased as

$$\frac{\partial \tilde{u}}{\partial \tilde{t}} = -\frac{\partial}{\partial \tilde{x}}(\tilde{v}_u \tilde{u}) + \tilde{s}_u, \tag{13}$$

and flux boundary conditions can be specified by prescribing boundary values for $\tilde{v}_u \tilde{u}$. Note that on the moving boundary $x = R(t)$, flux boundary conditions must be defined relative to the frame in which the boundary is stationary.

Macrophage foam cells undergo undirected random motion (45) and directed chemotactic motion towards modLDL. The latter is a simplification of the full process whereby modLDL stimulates the production of chemoattractants. Foam cells also undergo chemotactic motion towards dead material, in response to find-me signals expressed by the dying cells (33). ModLDL, HDL, and dead material are also assumed to undergo diffusive motion. For simplicity, we assign constant diffusion coefficients for these phases, and constant chemotactic coefficients for foam cells towards both modLDL and dead material. This gives the flux terms

$$\tilde{J}_m = -\tilde{D}_m \frac{\partial \tilde{m}}{\partial \tilde{x}} + \tilde{\chi}_l \tilde{m} \frac{\partial \tilde{l}}{\partial \tilde{x}} + \tilde{\chi}_c \tilde{m} \frac{\partial}{\partial \tilde{x}}(\tilde{c} + \tilde{b}), \tag{14}$$

$$\tilde{J}_u = -\tilde{D}_u \frac{\partial \tilde{u}}{\partial \tilde{x}} \quad \text{for } \tilde{u} = \tilde{c}, \tilde{b}, \tilde{l}, \tilde{h}, \tilde{p}, \tag{15}$$

Here, $\tilde{D}_m$, $\tilde{D}_c$, $\tilde{D}_l$, $\tilde{D}_h$, and $\tilde{D}_p$ are random motion or diffusion coefficients for macrophages, dead material, modLDL, saturated HDL, and unsaturated HDL respectively and $\tilde{\chi}_l$ and $\tilde{\chi}_c$ are chemotactic coefficients for foam cells in response to modLDL and dead material respectively. The dead cell phases share a diffusion constant, so $\tilde{D}_b = \tilde{D}_c$.

For the remaining phase $\tilde{a}$, we assume intracellular cholesterol droplets have the same phase velocity as the macrophages transporting them, giving the intracellular cholesterol flux term

$$\tilde{v}_a = \tilde{v}_m, \tag{16}$$

$$\text{or} \quad \tilde{J}_a = \frac{\tilde{a}}{\tilde{m}} \tilde{J}_m. \tag{17}$$

The endothelium allows a net influx of native LDL particles. Intimal LDL becomes chemically modified over much faster timescales than plaque growth timescales (18). We therefore model native LDL deposition as an boundary influx of modLDL. Nascent HDL particles also enter the plaque through the endothelium. For simplicity, we ignore HDL particles that already carry large amounts of cholesterol. HDL's antioxidant properties will reduce the rate of formation of pro-atherogenic modLDL by inhibiting LDL oxidation (9). We model this by allowing the modLDL influx to decrease as the endothelial HDL influx increases. This gives the following flux boundary conditions for the extracellular lipoprotein species at $\tilde{x} = 0$:

$$\tilde{v}_l \tilde{l} \Big|_{\tilde{x}=0} = \frac{\tilde{\sigma}_l}{1 + \tilde{\sigma}_p/\tilde{\sigma}_{p,\text{inh}}} \Big|_{\tilde{x}=0}, \tag{18}$$

$$\tilde{v}_h \tilde{h} \Big|_{\tilde{x}=0} = 0, \tag{19}$$

$$\tilde{v}_p \tilde{p} \Big|_{\tilde{x}=0} = \tilde{\sigma}_p, \tag{20}$$

$$\tag{21}$$

where $\tilde{\sigma}_l$ is the base rate of LDL deposition, $\tilde{\sigma}_p$ is the rate of unsaturated HDL deposition, and $\sigma_{p,\text{inh}}$ is the value of $\tilde{\sigma}_p$ required to inhibit LDL oxidation by a factor of 1/2.

Monocyte recruitment into the intima requires adhesion molecules expressed by endothelial cells and chemokines produced in the intima (30). These are expressed in response to the presence of modified LDL. For simplicity, we assume monocyte recruitment is proportional to the modLDL concentration at the endothelium. HDL's anti-inflammatory properties inhibit monocyte recruitment (4). We model this inhibition using the same functional form that we use for the inhibition of LDL oxidation. We also assume that monocyte differentiation timescales are much faster than those for plaque growth (6), and so we model monocyte recruitment as a macrophage boundary influx. The intracellular phase $a$ only includes cholesterol from ingested modLDL and not the endogenous cholesterol content of new cells, so it has an endothelial boundary flux of zero. Cells do not die until they are already inside the intima, so the dead material and intracellular cholesterol phases have zero flux endothelial boundary conditions. The flux boundary conditions for the intracellular phases at $\tilde{x} = 0$ are:

$$\tilde{v}_m \tilde{m} \Big|_{\tilde{x}=0} = \frac{\tilde{\sigma}_m \tilde{l}}{1 + \tilde{\sigma}_p/\tilde{\sigma}_{p,\text{inh}}} \Big|_{\tilde{x}=0}, \tag{22}$$

$$\tilde{v}_m \tilde{a} \Big|_{\tilde{x}=0} = 0, \tag{23}$$

$$\tilde{v}_c \tilde{c} \Big|_{\tilde{x}=0} = \tilde{v}_b \tilde{b} \Big|_{\tilde{x}=0} = 0, \tag{24}$$

where $\tilde{\sigma}_f$ is the rate of monocyte recruitment per unit of modLDL at the endothelium. Assuming $m$ is nonzero at the boundary, equation (12) implies that the boundary condition for $\tilde{a}$ can be simplified to

$$\tilde{a} \Big|_{\tilde{x}=0} = 0. \tag{25}$$

Foam cells are capable of emigrating into the lymphatic vasculature. HDL apolipoproteins promote foam cell emigration by upregulating receptors involved in emigration (21; 55).

Emigration is also inhibited in foam cells with large amounts of free cholesterol (42). We model these by allowing the medial foam cell efflux to decrease with the total lipid fraction $\bar{a} = \frac{\tilde{a}}{\tilde{m}+\tilde{a}}$, and to increase with total HDL levels, with eventual saturation. We also assume that both types of HDL particle can perfuse through the elastic media into the arterial lymphatics (29), at a rate proportional to their concentration. The remaining phases are given zero flux medial boundary conditions. The resulting boundary conditions at $\tilde{x} = \tilde{R}(\tilde{t})$ are

$$(\tilde{v}_m - \tilde{R}')\tilde{m}\Big|_{\tilde{x}=\tilde{R}} = \frac{\tilde{p}+\tilde{h}}{(\tilde{p}+\tilde{h})+\tilde{h}_{\text{egr}}}\tilde{\sigma}_e(1-\bar{a})\tilde{m}\Big|_{\tilde{x}=\tilde{R}}, \tag{26}$$

$$(\tilde{v}_h - \tilde{R}')\tilde{h}\Big|_{\tilde{x}=\tilde{R}} = \tilde{\sigma}_h \tilde{h}, \tag{27}$$

$$(\tilde{v}_p - \tilde{R}')\tilde{p}\Big|_{\tilde{x}=\tilde{R}} = \tilde{\sigma}_h \tilde{p}, \tag{28}$$

$$(\tilde{v}_u - \tilde{R}')\tilde{u}\Big|_{\tilde{x}=\tilde{R}} = 0 \quad \text{for } \tilde{u} = \tilde{l}, \tilde{c}, \tilde{b}, \tag{29}$$

where $\tilde{\sigma}_h$ is the average HDL particle efflux rate, $\tilde{\sigma}_e$ is the maximum uninhibited foam cell emigration velocity, and $h_{\text{egr}}$ governs the saturation of HDL-promoted emigration.

We note that the single boundary condition for $\tilde{a}$ at $\tilde{x} = 0$ is sufficient to achieve model closure. Since the total flux $\tilde{J}_a + \tilde{v}\tilde{a}$ contains no $\tilde{x}$ derivatives of $\tilde{a}$, the continuity equation (2) for phase $\tilde{a}$ is spatially first order, coupled to second order equations for the remaining species. The boundary flux of $\tilde{a}$ at $\tilde{x} = \tilde{R}(\tilde{t})$ can be derived directly from the boundary condition for $\tilde{m}$ and the fact that $\tilde{a}$ and $\tilde{m}$ share a single foam cell phase velocity (equations (17) and (26)). This flux is given by

$$(\tilde{v}_m - \tilde{R}')\tilde{a}\Big|_{\tilde{x}=\tilde{R}} = \frac{\tilde{p}+\tilde{h}}{(\tilde{p}+\tilde{h})+\tilde{h}_{\text{egr}}}\tilde{\sigma}_e(1-\bar{a})\tilde{a}\Big|_{\tilde{x}=\tilde{R}}. \tag{30}$$

**Initial conditions**

For initial conditions, we assume that the intima is populated with a small number of resident macrophages. We ignore any modLDL and HDL already present, as well as endogenous cellular cholesterol, so that macrophages have zero accumulated cholesterol prior to the initial endothelial injury. This gives us the initial conditions

$$\tilde{m}(\tilde{x}, 0) = N_0, \tag{31}$$

$$\tilde{u}(\tilde{x}, 0) = 0 \quad \text{for } \tilde{u} = \tilde{a}, \tilde{l}, \tilde{c}, \tilde{b}, \tilde{h}, \tilde{p}, \tag{32}$$

$$\tilde{R}(0) = \tilde{x}_S, \tag{33}$$

where $\tilde{x}_S$ is the diameter of a typical mouse macrophage.

**Model closure**

The last steps required to close the model are to define the mixture velocity $\tilde{v}(\tilde{x}, \tilde{t})$ and the velocity of the medial boundary $\tilde{R}'(\tilde{t})$. Models in more than one spatial dimension would require additional assumptions in order to determine $\tilde{v}(\tilde{x}, \tilde{t})$ and the boundary normal velocity, e.g. axial symmetry, or a velocity-pressure relation such as Darcy's law ($\tilde{v} = -k\nabla \tilde{P}$) (26; 36), along with additional constitutive assumptions and boundary conditions for the pressure $P$. Our model only considers motion along the radial direction, and so the existing continuity equations, boundary and initial conditions, and no-voids condition are sufficient to solve the system.

Summing over the phase continuity equations (2) for the volume-occupying phases and applying the no-voids condition (1) gives

$$0 = -\frac{\partial}{\partial \tilde{x}}\left(\left(1 + \frac{\tilde{a}}{\tilde{m}}\right)\left(-\tilde{D}_m\frac{\partial \tilde{m}}{\partial \tilde{x}} + \tilde{\chi}_l \tilde{m}\frac{\partial \tilde{l}}{\partial \tilde{x}} + \tilde{\chi}_c \tilde{m}\frac{\partial}{\partial \tilde{x}}(\tilde{c} + \tilde{b})\right)\right.$$
$$\left. - \tilde{D}_l\frac{\partial \tilde{l}}{\partial \tilde{x}} - \tilde{D}_c\frac{\partial}{\partial \tilde{x}}(\tilde{c} + \tilde{b}) - \tilde{D}_h\frac{\partial \tilde{h}}{\partial \tilde{x}} + N_0 \tilde{v}\right). \quad (34)$$

The mixture velocity is given by

$$\tilde{v} = \frac{1}{N_0}\left[\left(1 + \frac{\tilde{a}}{\tilde{m}}\right)\left(\tilde{D}_m\frac{\partial \tilde{m}}{\partial \tilde{x}} - \tilde{\chi}_l \tilde{m}\frac{\partial \tilde{l}}{\partial \tilde{x}} - \tilde{\chi}_c \tilde{m}\frac{\partial}{\partial \tilde{x}}(\tilde{c} + \tilde{b})\right)\right.$$
$$\left. + \tilde{D}_l\frac{\partial \tilde{l}}{\partial \tilde{x}} + \tilde{D}_c\frac{\partial}{\partial \tilde{x}}(\tilde{c} + \tilde{b}) + \tilde{D}_h\frac{\partial \tilde{h}}{\partial \tilde{x}} + \frac{\tilde{\sigma}_m \tilde{l} + \tilde{\sigma}_l}{1 + \tilde{\sigma}_p/\tilde{\sigma}_{p,\text{inh}}}\bigg|_{\tilde{x}=0}\right], \quad (35)$$

and the domain grows with rate

$$\frac{d\tilde{R}}{d\tilde{t}} = \frac{1}{N_0}\left[\frac{\tilde{\sigma}_m \tilde{l} + \tilde{\sigma}_l}{1 + \tilde{\sigma}_p/\tilde{\sigma}_{p,\text{inh}}}\bigg|_{\tilde{x}=0} - \frac{\tilde{p} + \tilde{h}}{(\tilde{p} + \tilde{h}) + \tilde{h}_{\text{egr}}}\tilde{\sigma}_e(1 - \tilde{a})(\tilde{m} + \tilde{a})\bigg|_{\tilde{x}=\tilde{R}} - \tilde{\sigma}_h \tilde{h}\bigg|_{\tilde{x}=\tilde{R}}\right]. \quad (36)$$

## 2.1 Nondimensionalisation and parameter estimates

We nondimensionalise the model by applying the rescalings

$$u = \frac{\tilde{u}}{N_0}, \quad t = \frac{\tilde{t}}{t_S}, \quad x = \frac{\tilde{x}}{x_S}, \quad (37)$$

where the species density rescaling applies to all species $u = m, c, l, a, b, h, p$. Here, $t_S$ and $x_S$ are some reference timescale and length scale, and $N_0$ is the total phase density.

Plaque studies in ApoE knockout mice are typically carried out over periods of weeks to months (65; 49). We choose a characteristic timescale of about one week, setting $t_S = 6 \times 10^5$ s. Lengths are rescaled by the diameter of a typical mouse macrophage. Macrophages in mice are observed to vary in size from under 14 µm for murine bone marrow-derived macrophages (11) to 20 µm for murine peritoneal macrophages (44). We choose a characteristic size of 16 µm. Since our model is only in one spatial dimension, certain parameter estimates require the conversion of three-dimensional volume densities to one-dimensional linear densities. For the purposes of parameter estimation, we consider a cylindrical section through the intima whose circular cross-section is transverse to the positive $x$ vector and has diameter $x_S$, so that the maximum linear phase density $N_0$ is one cell (or equivalent cell volume) per unit of $x_S$.

Most model parameters carry over from the three-phase model in Ahmed et al (1), and these parameters are given the same or similar values. Below, we focus on parameters that are new to the six-phase HDL model.

The HDL degradation parameter $\tilde{\delta}_p$ is based on *in vitro* studies (28; 34) where HDL was incubated with highly oxidative copper ions. Lipid oxidative by-products were observed in significant quantities after about 12 hours. The reverse cholesterol transport parameter $\tilde{\mu}_h$ is estimated from an *in vitro* study (47) where mouse peritoneal macrophages were incubated with modified LDL to produce foam cells, and then incubated for 24 hours with HDL. About 25 % to 40 % of the macrophages' cholesterol content was incorporated by HDL, and we choose a typical timescale of 16 to 24 hours.

The remaining parameters are given order of magnitude estimates that produce biologically reasonable results. The HDL diffusion coefficients $\tilde{D}_p$ and $\tilde{D}_h$ are given similar but greater values than for LDL, since HDL particles are smaller than LDL particles (22), and particle size is known to limit the transfer rate of lipoproteins across endothelial and membrane barriers (29). The range of values for the unsaturated HDL influx $\tilde{\sigma}_p$ is chosen to cover a similar range to the LDL influx $\tilde{\sigma}_l$, where $\tilde{\sigma}_{p,\text{inh}}$ is situated in the upper range of these values. The medial HDL clearance rate $\tilde{\sigma}_h$ is chosen so that the HDL efflux $\tilde{\sigma}_h \tilde{h}|_{\tilde{x}=\tilde{R}}$ and maximum base foam cell emigration flux $\tilde{\sigma}_e(\tilde{m}+\tilde{a})|_{\tilde{x}=\tilde{R}}$ are similar in magnitude. The base foam cell emigration velocity here was chosen based on values of $\tilde{\sigma}_f$ that were found to enable plaque stabilisation in Ahmed et al (1). Test model runs were used to estimate typical values for $h|_{x=R}$, with $h_{\text{egr}}$ chosen to be characteristic of these values.

Tables 1 and 2 summarise the model parameter values before and after rescaling. The full nondimensional model is given in Appendix A.

## 2.2 Numerical solution

The full system was solved numerically using the method of lines. We first used the transformation $(x, t) \rightarrow (y, \tau) = (x/R(t), t)$ to map the nondimensional system onto a fixed spatial domain $[0, 1]$. giving a system of parabolic PDEs coupled to an ODE for the domain length $R(\tau)$. The transformed system was then reduced to a set of time-dependent ODEs by discretising spatial derivatives using a central differencing approximation. The resulting ODEs were solved using a backward differentiation formula method in Python using SciPy's integration libraries.

# 3 HDL-free plaque growth

We first look at plaque growth in the absence of HDL, and use some illustrative examples to explore how the rates of cell death, inflammation (i.e. monocyte recruitment), efferocytosis, and emigration influence plaque structure and growth. We set the unsaturated HDL influx to $\sigma_p = 0$ in this section.

Table 1: Dimensional model parameter estimates

| Parameter + Description | | Dimensional estimate | Sources |
|---|---|---|---|
| $t_S$ | reference time scale | $6 \times 10^5\,\text{s}$ | — |
| $x_S$ | reference length scale | $1.6 \times 10^{-5}\,\text{m}$ | (11) |
| $N_0$ | maximum phase density | $1\,\text{cell}/x_S$ | — |
| $\tilde{D}_m$ | macrophage random motility coefficient | $\ll 10^{-14}\,\text{m}^2\,\text{s}^{-1}$ | (45) |
| $\tilde{D}_l$ | modLDL diffusion coefficient | $> \tilde{D}_m$ | order of mag est |
| $\tilde{D}_c$ | apoptotic body diffusion coefficient | $< \tilde{D}_m$ | order of mag est |
| $\tilde{D}_h$ | saturated HDL diffusion coefficient | $> \tilde{D}_l$ | order of mag est |
| $\tilde{D}_p$ | unsaturated HDL diffusion coefficient | $> \tilde{D}_l$ | order of mag est |
| $\tilde{\chi}_l$ | macrophage chemotactic coefficient (modLDL) | $\gg \tilde{D}_m/N_0$ | order of mag est, (45) |
| $\tilde{\chi}_c$ | macrophage chemotactic coefficient (apoptotic bodies) | $\sim \tilde{\chi}_l$ | order of mag est |
| $\tilde{\mu}_a$ | macrophage apoptosis rate | $\sim 0.25\,\text{h}^{-1}$ | (66) |
| $\tilde{\mu}_p$ | modLDL phagocytosis rate | $\sim (30\,\text{min})^{-1}/N_0$ | (70) |
| $\tilde{\mu}_e$ | efferocytosis rate | $\sim \tilde{\mu}_a/N_0$ | order of mag est, (66) |
| $\tilde{\mu}_h$ | reverse cholesterol transport efflux rate | $\sim \tilde{\mu}_a N_0$ | (47) |
| $\tilde{\delta}_p$ | HDL degradation rate | $\sim (12\,\text{h})^{-1}$ | (28) |
| $\tilde{\sigma}_m$ | monocyte recruitment rate per unit modLDL | $\sim 5 \times 10^5\,\text{cell}\,\text{m}^{-2}\,\text{s}^{-1} \cdot x_S^2/N_0$ | (31) |
| $\tilde{\sigma}_l$ | LDL deposition rate | $\sim \tilde{\sigma}_m N_0$ | order of mag est |
| $\tilde{\sigma}_p$ | unsaturated HDL deposition rate | $\sim \tilde{\sigma}_l$ | order of mag est |
| $\tilde{\sigma}_{p,\text{inh}}$ | HDL inhibition of LDL oxidation and monocyte recruitment | $\sim \sigma_p$ | order of mag est |
| $\tilde{\sigma}_e$ | uninhibited foam cell emigration velocity | $\sim \tilde{\sigma}_m, \tilde{\sigma}_l/N_0$ | order of mag est |
| $\tilde{h}_{\text{egr}}$ | saturation of HDL-upregulated emigration | $< N_0$ | order of mag est |
| $\tilde{\sigma}_h$ | HDL particle efflux rate | $> \tilde{\sigma}_e$ | order of mag est |

## 3.1 Structure and composition without emigration

We focus on non-emigratory cell behaviour in this section, and we fix the emigration rate to $\sigma_e = 0$. We also fix the LDL influx to $\sigma_l = 10$.

Figure 2(a) illustrates the plaque's spatial structure for a base case scenario, where the rate of efferocytosis is low compared to cell death ($\sigma_m = 100$, $\mu_e = 30$). Live macrophages are only found close to the endothelium where they are recruited, along with some newly deposited modLDL that has yet to be ingested. The middle and deep plaque tissue consists solely of dead cellular material with no live cells, and the rate at which dead material is efferocytosed here is zero. From Figure 3(a), the intima quickly approaches a state where the rate of growth is constant. The endothelial monocyte recruitment rate settles to a constant value (Figure 3(b)), as does the ratio between the monocyte and modLDL influxes, and the rate at which dead material accumulates. (Figure 3(c)).

Complete cell death in the deep plaque can be prevented by sufficiently increasing the ratio of efferocytosis to cell death. Figure 2(b) and (c) depict two plaques where this has been achieved by different means. The scenario in Figure 2(b) increases the efferocytic capacity parameter $\mu_e$ to more than double the base case value ($\sigma_m = 100$, $\mu_e = 80$). The higher efferocytic capacity means foam cells are better able to clear dead cells, thus lowering the proportion of dead cells. Due to the higher amount of live cells near the endothelium available to ingest pro-inflammatory modLDL, monocyte recruitment also decreases slightly. Average intracellular cholesterol loads are higher as a result, which causes slightly higher cell death rates $\mu_a \bar{a}$, but this is more than compensated for by the much higher rates of efferocytic uptake.

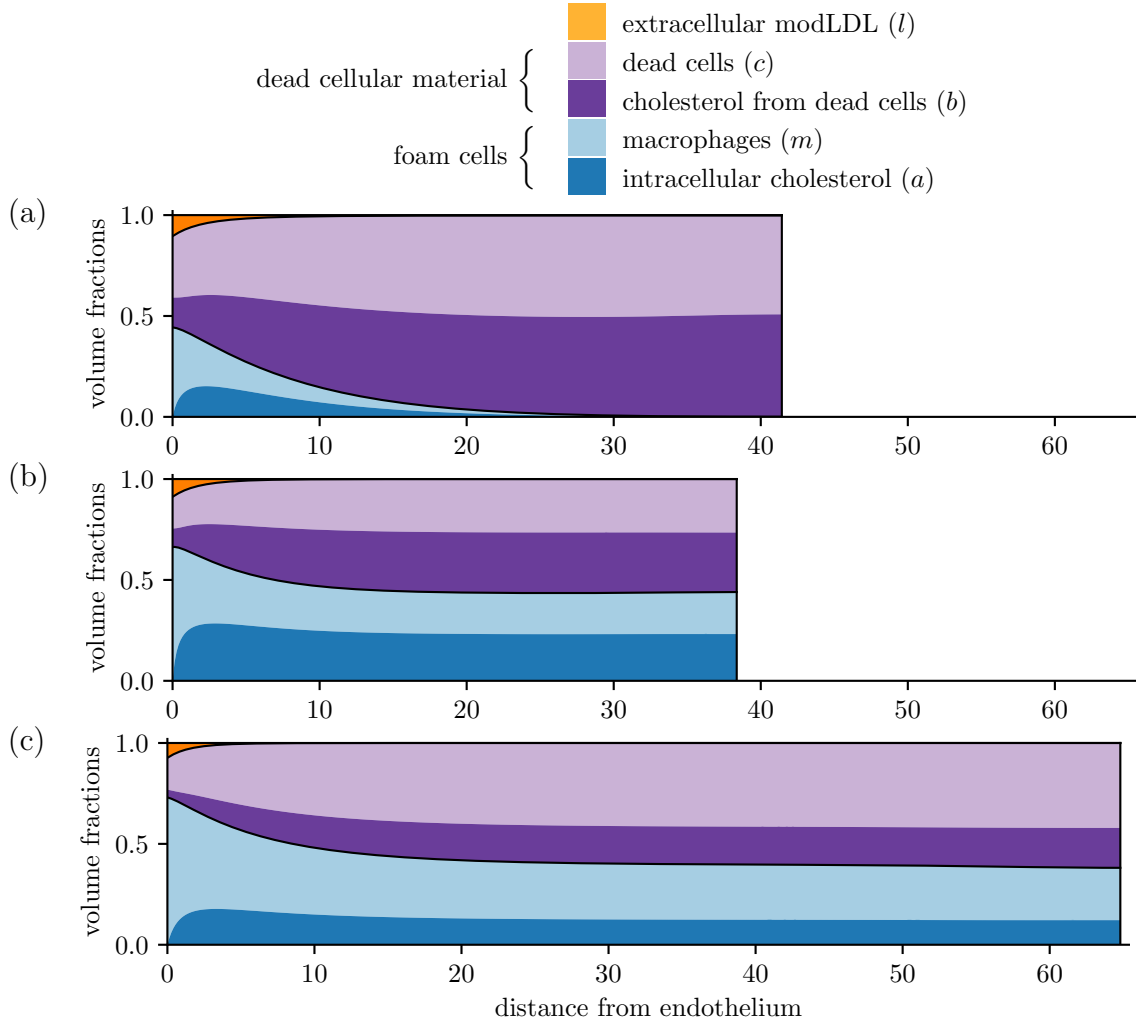

Figure 2: Plaque structure plots at $t = 2$ depicting solutions to the HDL-free model with $\sigma_l = 10$ and $\sigma_e = 0$. Plots depict (a) a base scenario with $\sigma_m = 100$ and $\mu_e = 30$ that sees complete cell die-off in the deep plaque, (b) a higher efferocytosis scenario with $\sigma_m = 100$ and $\mu_e = 80$, and (c) a higher inflammation scenario with $\sigma_m = 300$ and $\mu_e = 30$.

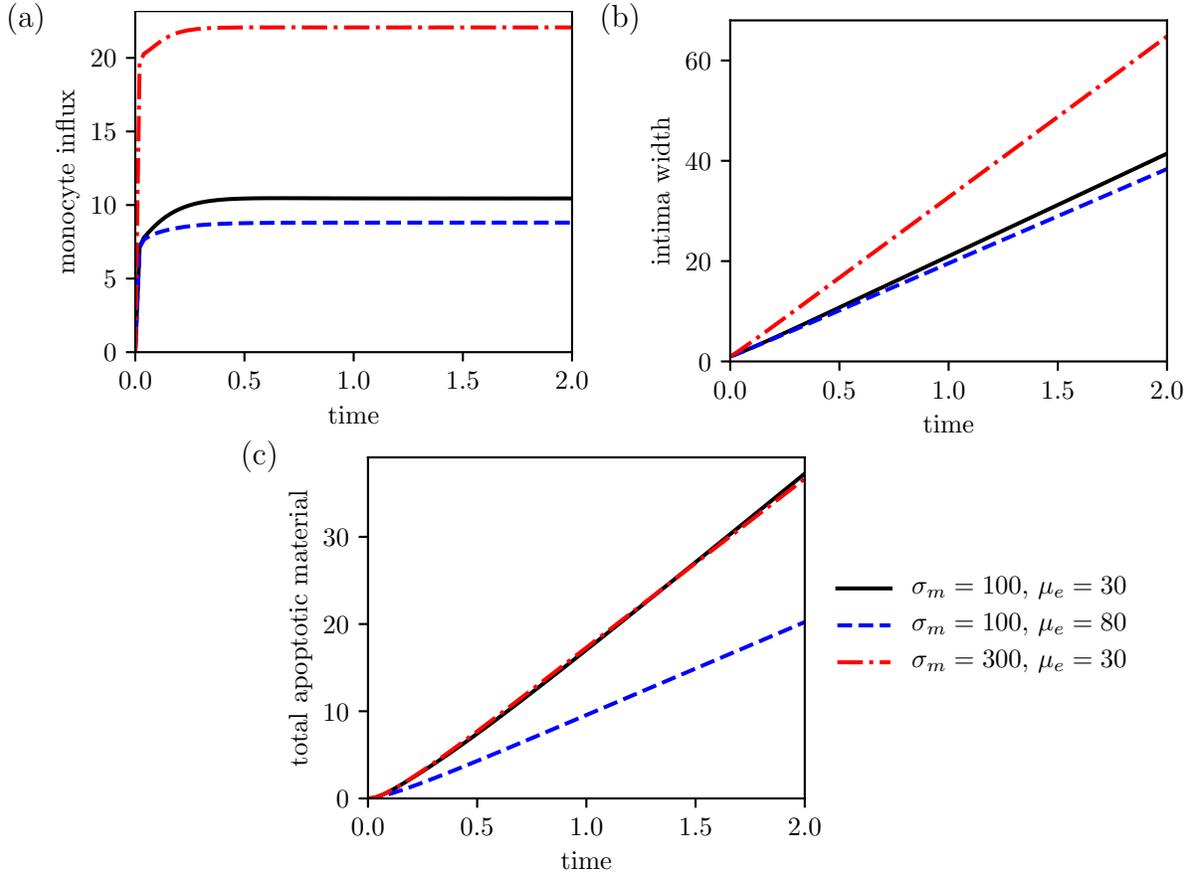

Figure 3: Comparison of (a) the endothelial monocyte influx $\sigma_m l|_{x=0}$, (b) the intima width $R$, and (c) the total apoptotic material $\int_0^R (c+b)\,\mathrm{d}x$ as a function of time for the three plaque scenarios in Figure 2.

Table 2: Rescaled parameters for the nondimensionalised model

| Parameter | Rescaling | Nondimensional estimate |
|---|---|---|
| $D_m$ | $\tilde{D}_f/(x_S^2/t_S)$ | 10 |
| $D_l$ | $\tilde{D}_l/(x_S^2/t_S)$ | 50 |
| $D_c$ | $\tilde{D}_c/(x_S^2/t_S)$ | 5 |
| $D_h$ | $\tilde{D}_h/(x_S^2/t_S)$ | 80 |
| $D_p$ | $\tilde{D}_p/(x_S^2/t_S)$ | 80 |
| $\chi_l$ | $\tilde{\chi}_l/(x_S^2/N_0 t_S)$ | 200 |
| $\chi_c$ | $\tilde{\chi}_c/(x_S^2/N_0 t_S)$ | 200 |
| $\mu_a$ | $\tilde{\mu}_a/(1/t_S)$ | 40 |
| $\mu_p$ | $\tilde{\mu}_p/(1/N_0 t_S)$ | 150 |
| $\mu_e$ | $\tilde{\mu}_e/(1/N_0 t_S)$ | 10 to 100 (variable) |
| $\mu_h$ | $\tilde{\mu}_h/(1/N_0 t_S)$ | 10 |
| $\delta_p$ | $\tilde{\delta}_p/(1/t_S)$ | 15 |
| $\sigma_m$ | $\tilde{\sigma}_f/(x_S/t_S)$ | 50 to 500 (variable) |
| $\sigma_l$ | $\tilde{\sigma}_l/(N_0 x_S/t_S)$ | 10 to 50 (variable) |
| $\sigma_p$ | $\tilde{\sigma}_p/(N_0 x_S/t_S)$ | 10 to 100 (variable) |
| $\sigma_{p,\text{inh}}$ | $\tilde{\sigma}_{p,\text{inh}}/(N_0 x_S/t_S)$ | 80 |
| $\sigma_e$ | $\tilde{\sigma}_e/(x_S/t_S)$ | 100 |
| $h_{\text{egr}}$ | $\tilde{h}_{\text{egr}}/N_0$ | 0.01 |
| $\sigma_h$ | $\tilde{\sigma}_h/(x_S/t_S)$ | 200 |

The third scenario in Figure 2(c) instead considers a plaque with higher inflammation, where the monocyte recruitment parameter $\sigma_m$ has been increased ($\sigma_m = 300$, $\mu_e = 30$). Here, the higher monocyte influx also results in a higher ratio of incoming cellular material to cholesterol. The cytotoxic cholesterol loads are distributed over more macrophages, thereby reducing the rates of cell death. Although the efferocytic capacity $\mu_e$ is identical, the lower death rates enable live foam cells to coexist with dead material. The increased proportion of live cells also means higher rates of efferocytic uptake of dead cells. Figure 4 summarises the medial foam cell density for various values of $\sigma_m$ and $\mu_e$, and shows consistent behaviour, where higher $\sigma_m$ or $\mu_e$ leads to a higher proportion of live cells.

While increasing either $\sigma_m$ or $\mu_e$ helps boost the proportion of live foam cells, the two parameters govern different aspects of plaque growth. For the higher efferocytosis (higher $\mu_e$) case, although $\mu_e$ does not directly govern any boundary fluxes, the increase in live cell proportions leads to greater consumption of modLDL near the endothelium. The lowered modLDL results in a slight decrease in monocyte recruitment, which slightly slows down plaque growth (Figure 2(b) and Figure 3(b)), at least in this case where emigration is zero.

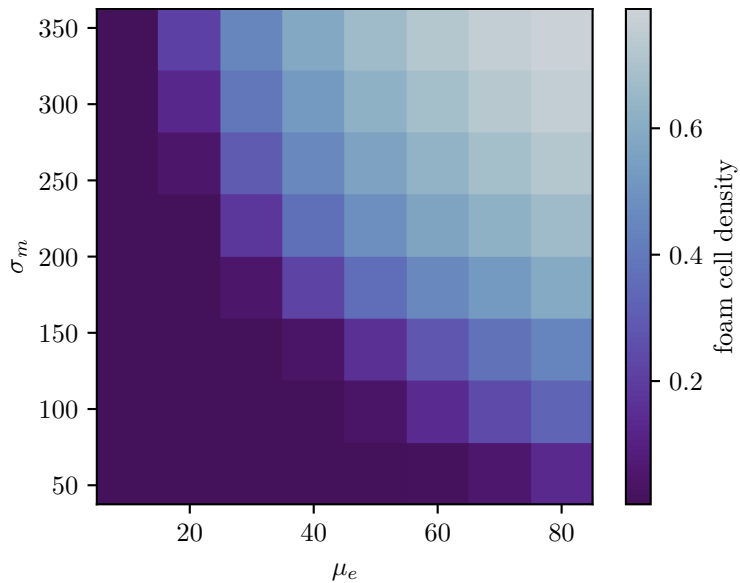

Figure 4: Heat map of the medial foam cell density $f|_{x=R}= (m + a)|_{x=R}$ for varying $\mu_e$ and $\sigma_m$. Values are for the HDL-free and emigration-free model ($\sigma_p$, $\sigma_e = 0$), with $\sigma_l = 10$, and measured at $t = 2$.

For the higher inflammation (higher $\mu_e$) case, increasing monocyte recruitment directly increases plaque growth due to the greater amounts of incoming cell material (Figure 2(c) and Figure 3(b)). The higher $\mu_e$ case accumulates new dead material at a much lower rate than the base case (Figure 3(c)), but not the high $\sigma_m$ case. Although the second and third cases both contain lower proportions of dead cells, this is offset in the high $\sigma_m$ case by the higher rate at which new cellular material is introduced.

## 3.2 Foam cell emigration

Foam cell emigration is key to slowing plaque growth or stabilising plaque size completely, because it provides a means for material to exit the intima. The degree to which emigration can slow plaque growth depends on the availability of live foam cells near the medial boundary.

Emigration is aided by high rates of efferocytic uptake, as we observed the three-phase model in Ahmed et al (1). In plaques with poor efferocytosis, emigration is negligible since all foam cells in the deep plaque die off, from Figure 5. This is true even when the base emigration velocity $\sigma_e$ is high. Scenarios with higher efferocytic capacity $\mu_e$ have lower rates of intimal growth, since the increased proportion of live foam cells leads to higher rates of emigration. For plaques with a higher base emigration velocity $\sigma_e$, sufficiently increasing $\mu_e$ allows plaques to settle to a stable size.

The effect of inflammation on emigration is less straightforward. Increasing $\sigma_m$ may either slow or accelerate plaque growth, depending on whether the increase in emigration is enough to counteract the increased monocyte influx. For plaques with heavily impaired efferocytosis such as in Figure 6, foam cell densities in the deep plaque are negligible. Increasing $\sigma_m$ here only increases the influx of new macrophages that will ultimately die and remain in the plaque. Inflammation therefore promotes plaque growth when efferocytic clearance is insufficient to prevent complete cell death.

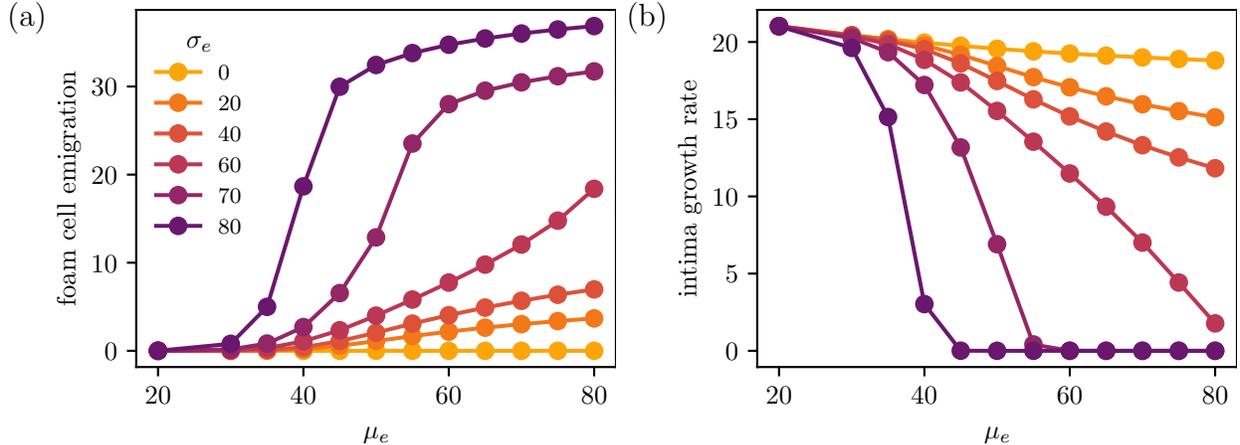

Figure 5: Comparison of (a) the foam cell emigration flux $(j_m + j_a)|_{x=R}$ and (b) the plaque growth rate $\frac{dR}{dt}$ as a function of the efferocytosis parameter $\mu_e$ for varying base egress velocities $\sigma_e$, where $\sigma_m = 100$, $\sigma_l = 10$, and $\sigma_p = 0$, measured at $t = 2$.

For scenarios with moderate to high efferocytosis, additional inflammation may encourage emigration. In Figure 7, increasing the monocyte influx reduces death rates by diluting the average foam cell cholesterol load. This has the dual effect of increasing deep plaque foam cell counts, and reducing the inhibitive effect of cholesterol loading on emigration. In some cases, the boost in emigration is outweighed by the increase in monocyte recruitment, and higher $\sigma_m$ has the net effect of promoting plaque growth. This is observed in Figure 7 for $\sigma_e = 10$. When $\sigma_e$ is higher, however, the diluted intracellular cholesterol loads increases emigration enough to outweigh the greater monocyte recruitment, thereby resulting in slower plaque growth, and even plaque stabilisation for the $\sigma_e = 60$ case.

## 4 Reverse cholesterol transport

In this section, we focus on how HDL-mediated reverse cholesterol transport (RCT) helps to reduce plaque growth. We consider the full model from Section 2 in Section 5, but this section, we only consider HDL's cholesterol efflux capabilities and its clearance from the plaque, and we ignore its anti-inflammatory and antioxidant properties and its upregulation of foam cell emigration. We simplify the macrophage and modLDL boundary conditions to become HDL-independent, so at the endothelial boundary,

$$v_m m \bigg|_{x=0} = \sigma_m l \bigg|_{x=0}, \quad v_l l \bigg|_{x=0} = \sigma_l, \tag{38}$$

and at the medial boundary,

$$v_m m \bigg|_{x=R} = \sigma_e(1-\overline{a}) m \bigg|_{x=R}, \quad v_m a \bigg|_{x=R} = \sigma_e(1-\overline{a}) a \bigg|_{x=R}. \tag{39}$$

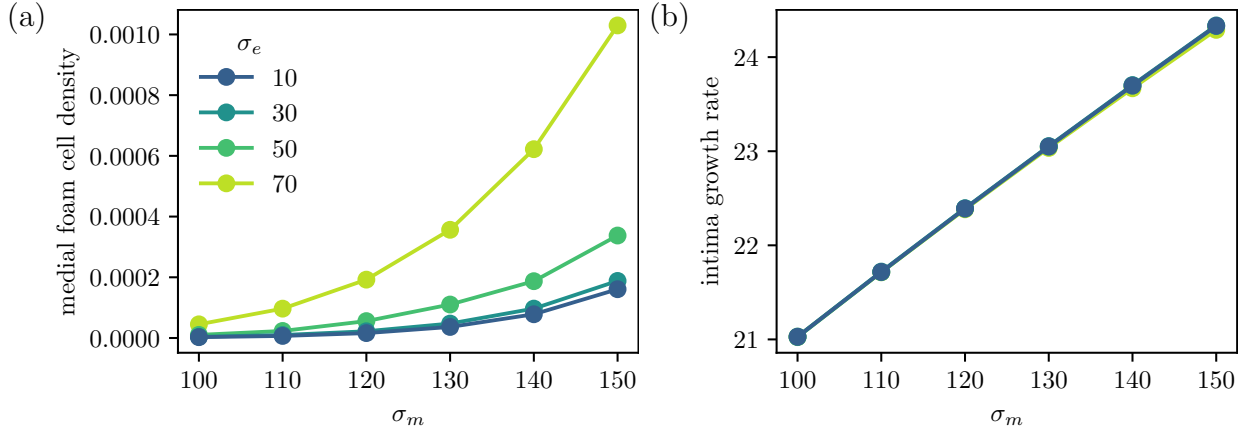

Figure 6: Comparison of (a) the medial foam cell density $(m+a)|_{x=R}$ and (b) the plaque growth rate $\frac{dR}{dt}$ as a function of the efferocytosis parameter $\sigma_m$ for base egress velocities $\sigma_e = 10 - 70$. Plots are for a scenario with poor efferocytic efficiency and few live foam cells, where $\mu_e = 20$, $\sigma_l = 10$, and $\sigma_p = 0$, measured at $t = 2$. Note that in (b), the plots for different $\sigma_e$ are indistinguishable.

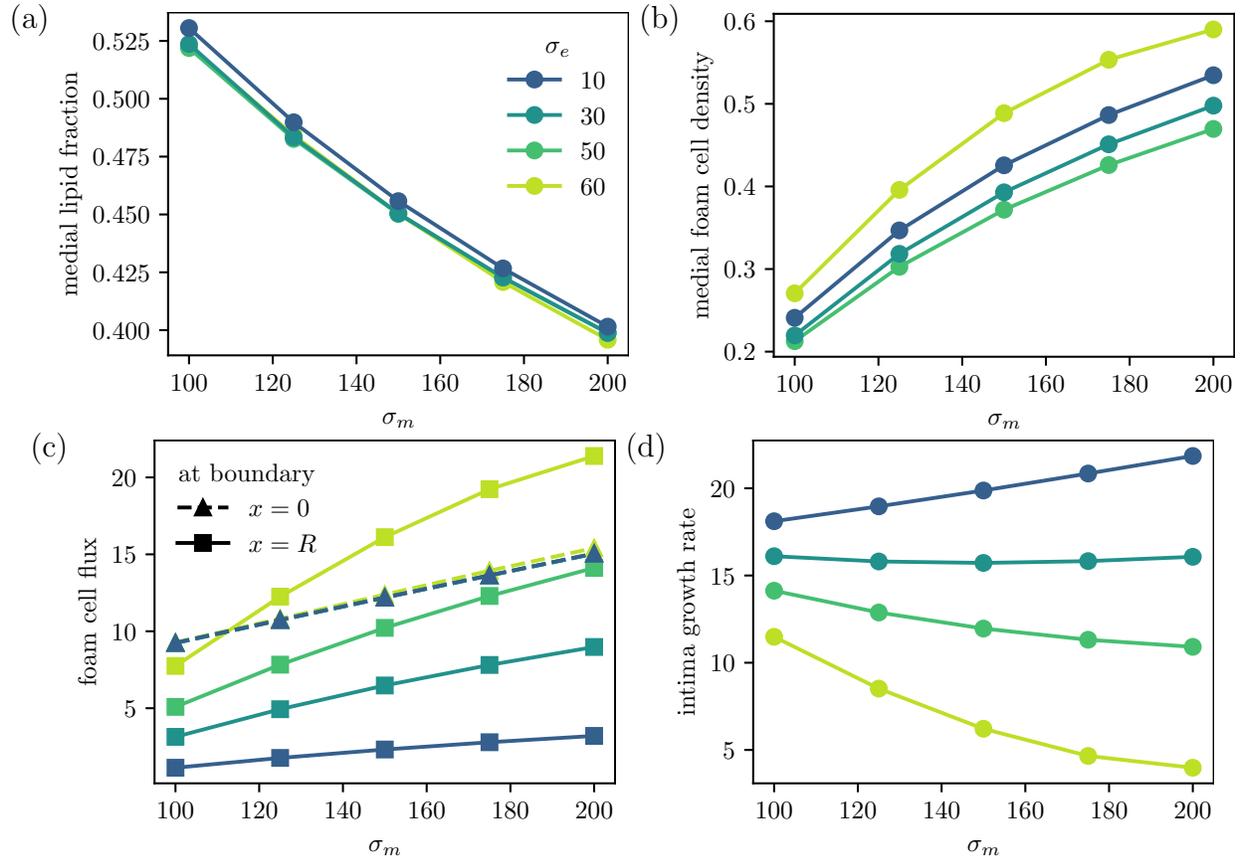

Figure 7: Comparison of (a) the medial foam cell cholesterol proportion $\frac{a}{m+a}|_{x=R}$, (b) the medial foam cell density $(m+a)|_{x=R}$, (c) the boundary foam cell fluxes $j_m + j_a$ at $x = 0, R$, and (d) the plaque growth rate $\frac{dR}{dt}$ as a function of the efferocytosis parameter $\sigma_m$, for base egress velocities $\sigma_e = 10 - 60$. Plots are for a scenario with moderate efferocytic efficiency, where $\mu_e = 60$, $\sigma_l = 10$, and $\sigma_p = 0$, measured at $t = 2$.

We note that in this model and the full model, since cellular material $m + c$ has no net source or sink terms and saturated HDL $h$ has only source terms, their net quantities over the plaque will increase in time unless there is a boundary efflux term to allow their removal. Both foam cell emigration and HDL clearance are therefore necessary for the existence of steady state solutions.

## 4.1 RCT vs HDL influx

The removal of cholesterol from foam cells by HDL has an unambiguously beneficial effect on reducing plaque growth. Figure 8(a) shows an unhealthy plaque, where foam cells in the deep plaque die off rapidly to create a core of dead material near the medial boundary. In Figure 8(b) and (c), introducing HDL prevents foam cell die-off. From Figure 9(a), higher influxes of unsaturated HDL enable higher rates of cholesterol efflux from foam cells. The resulting decrease in cytotoxic cholesterol loads causes a reduction in death rates, thereby boosting live foam cell quantities, as is observed in Figure 8.

The action of HDL is not spatially uniform, and is concentrated mostly near the endothelium. In Figure 8(d), the concentration of unsaturated HDL particles is highest at the endothelium, and decreases further into the plaque due to chemical degradation and conversion into saturated HDL. From Figure 9(a), cholesterol efflux activity is localised near but not at the endothelium, as foam cells nearest to the endothelium have yet to accumulate substantial amounts of cholesterol that can be offloaded onto unsaturated HDL particles. Figure 8(b) and (c) show that saturated HDL particles are present throughout the plaque due to diffusion and advection, but are still found in higher quantities close to the endothelium where most cholesterol efflux activity is taking place.

Figure 8 and Figure 9(b) show that the increase in RCT also has the effect of slowing plaque growth. From Figure 9(c), this is due to increased rates of both foam cell emigration and HDL clearance. Here, increased foam cell emigration is a consequence of lowered cholesterol loads, which both increase the numbers of live foam cells and reduce the inhibition of emigration due to cholesterol toxicity. From Figure 9(b), the rate at which new dead material accumulates is also slowed due to the combined effects of increased foam cell emigration and reduced cell death. In some cases, increasing the HDL influx $\sigma_p$ sufficiently can eventually stabilise plaque growth, as we observe in Figure 10. For fixed base foam cell emigration velocities $\sigma_e$, higher $\sigma_e$ allows plaque stabilisation for lower levels of HDL. We note that this is observed even with the simplified RCT-only HDL model (equations (38) and (39)). In the full model (Section 5), which accounts for HDL's anti-inflammatory and antioxidant properties, increasing HDL is expected to cause more dramatic decreases in the rates of plaque growth and dead material accumulation.

## 4.2 Efferocytosis, inflammation, and RCT

Cholesterol efflux rates are also influenced by efferocytosis and inflammation. Figure 11 shows that for a fixed HDL influx rate $\sigma_p$, increasing the efferocytosis rate $\mu_e$ enhances RCT by increasing live foam cell counts, since the cholesterol content of dead cells is inaccessible to HDL particles.

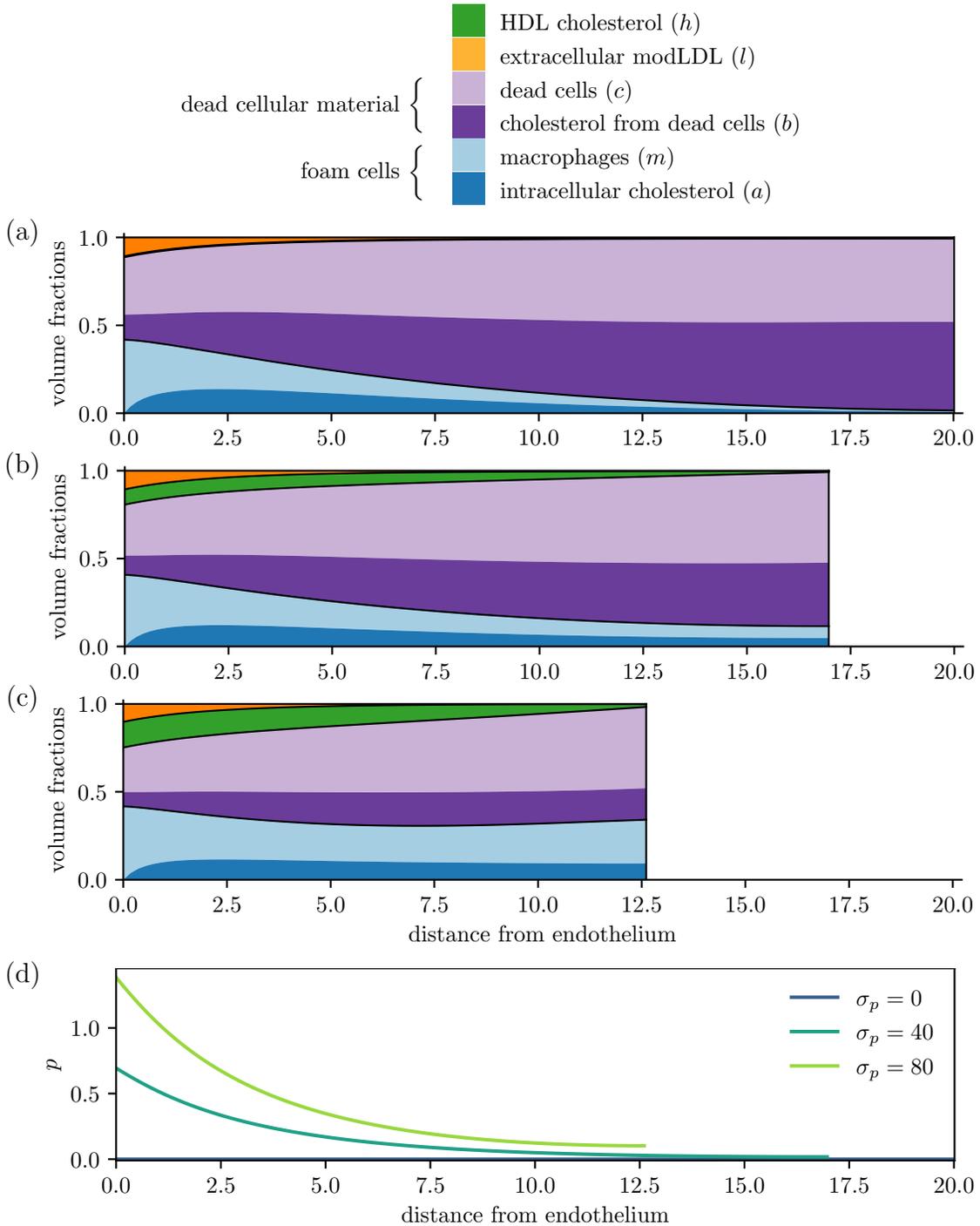

Figure 8: Plaque structure plots at $t = 1$ for varying unsaturated HDL influxes: (a) $\sigma_p = 0$, (b) $\sigma_p = 40$, and (c) $\sigma_p = 80$, with unsaturated HDL density $p$ plotted in (d). Solutions are for the simplified RCT-only HDL model (equations (38) and (39)) with $\sigma_m = 100$, $\sigma_l = 10$, $\sigma_e = 10$, and $\mu_e = 25$.

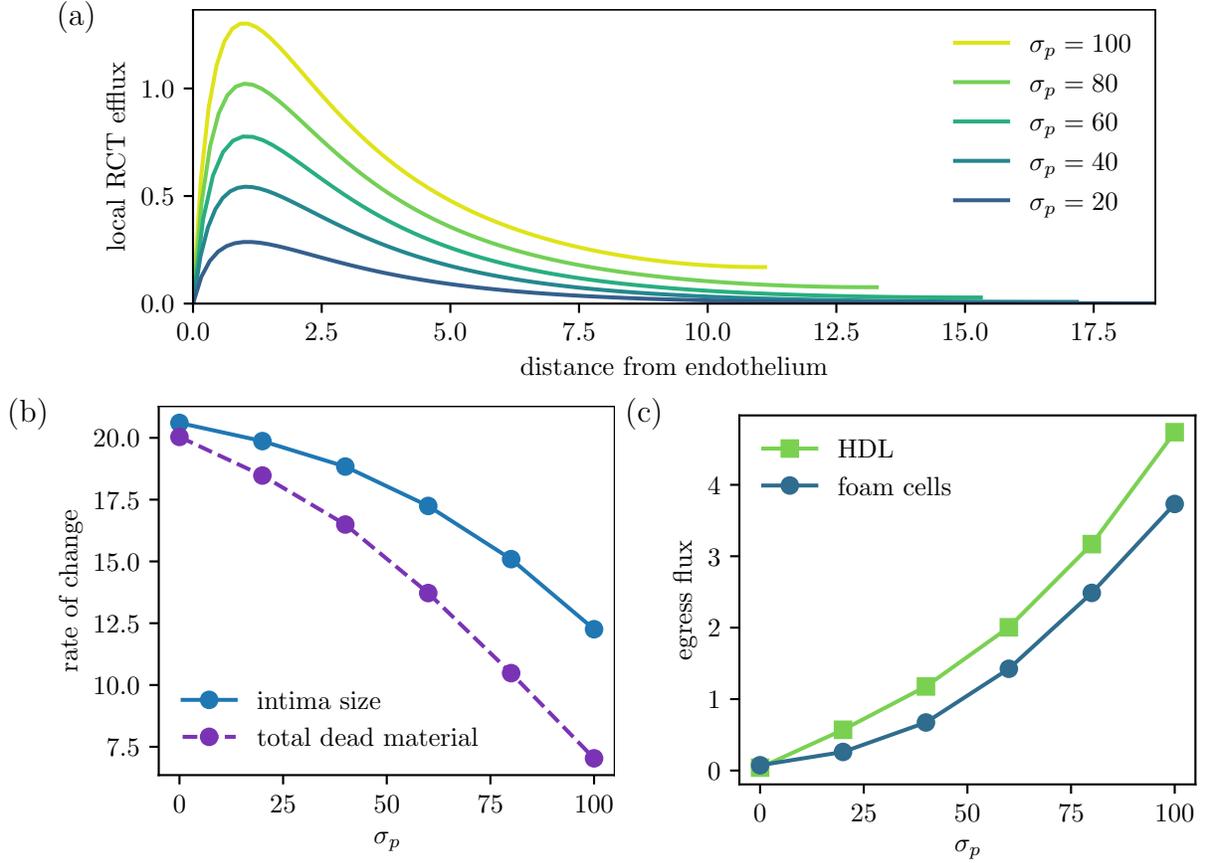

Figure 9: Comparisons of (a) the local cholesterol efflux rate from foam cells to HDL $\mu_h p a$, (b) the rates of change of the total plaque size $\frac{dR}{dt}$ and total dead material $\frac{d}{dt}\int_0^R (c+b)\,dx$, and (c) the medial HDL and foam cell effluxes $v_h h|_{x=R}$ and $v_m(m+a)|_{x=R}$, for varying HDL influxes. Solutions are for the simplified RCT-only HDL model (equations (38) and (39)) with $\sigma_m = 100$, $\sigma_l = 10$, $\sigma_e = 10$, and $\mu_e = 25$, plotted at $t=1$.

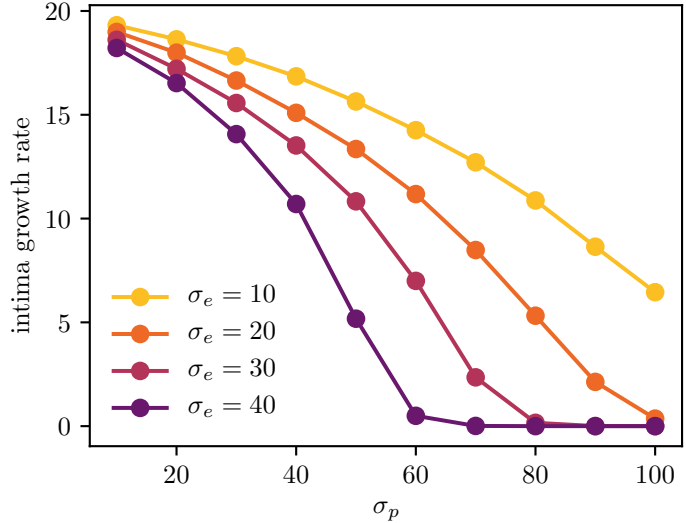

Figure 10: Comparison of the intima growth rate $\frac{dR}{dt}$ at $t = 2$ for varying HDL influx rates $\sigma_p$, plotted for different base foam cell emigration velocities $\sigma_e$. Solutions are for the simplified RCT-only HDL model (equations (38) and (39)) with $\sigma_m = 100$, $\sigma_l = 10$, and $\mu_e = 40$.

Increasing inflammation via $\sigma_m$ can either help or hinder RCT. Figure 12(a) and (c) depict a scenario with defective efferocytosis, where increasing inflammation boosts live foam cell counts by reducing average cholesterol loads. Here, higher amounts of cholesterol become accessible to HDL, thus improving cholesterol efflux. Figure 12(b) and (d) on the other hand depict a plaque with healthier efferocytosis and higher live foam cell counts. Here, increasing inflammation only dilutes the already relatively low average cholesterol loads. As unsaturated HDL is concentrated mostly near the endothelium, more of this cholesterol becomes spatially inaccessible, resulting in lower rates of cholesterol efflux when inflammation is increased.

### 4.3 LDL and RCT

LDL is the sole source of cholesterol in our model, and increasing the influx of LDL into the intima counteracts the positive effects of reverse cholesterol transport.

From Figure 13(a), higher LDL influxes $\sigma_l$ cause the endothelial monocyte influx to increase in response to the higher densities of modLDL near the endothelium (not shown). The proportional increase in monocyte recruitment is smaller than the proportional increase in incoming cholesterol, and higher $\sigma_l$ results in a higher average cholesterol load per macrophage (Figure 13(b)), both with or without cholesterol removal by HDL. This leads to higher foam cell death rates and increased inhibition of emigration, which results in lower rates of foam cell emigration at the medial boundary (Figure 13(c)). The increased foam cell cholesterol loads lead to higher overall rates of cholesterol efflux onto unsaturated HDL particles (not shown), but HDL clearance at the medial boundary decreases (Figure 13(c)). This is because the larger plaque size means saturated HDL is distributed over a larger space, and less of it is available near the medial boundary. The net effect of more lipid and cellular material entering the plaque and less material exiting is that higher $\sigma_l$ accelerates plaque

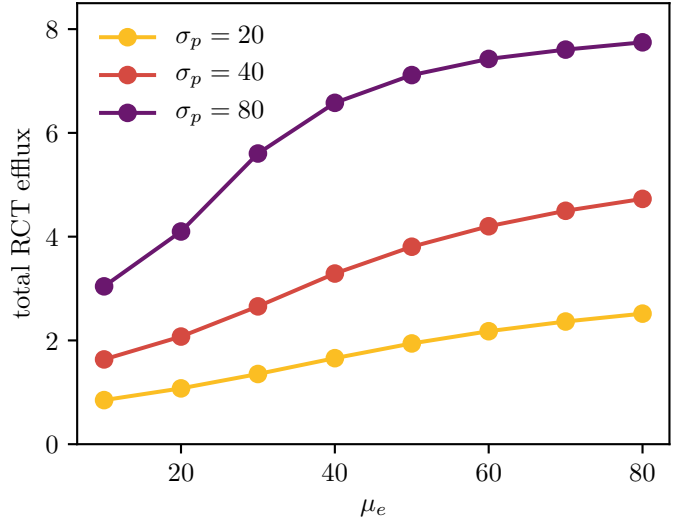

Figure 11: Comparison of the total cholesterol efflux rate $\int_0^R \mu_h p a \, dx$ for varying efferocytosis rates $\mu_e$, plotted for several different HDL influxes $\sigma_p$. Solutions are for the simplified RCT-only HDL model (equations (38) and (39)) with no foam cell emigration ($\sigma_e = 0$, $\sigma_m = 100$, $\sigma_l = 10$), plotted at $t = 1$.

growth (Figure 13(d)).

## 5 The LDL-HDL lipid profile

In this section, we explore how the levels of both LDL and HDL in the bloodstream influence plaque development using the full model from Section 2. We consider scenarios with moderate inflammation ($\sigma_m = 100$) and poor efferocytosis ($\mu_e = 30$) to start with, but we later consider the effects of changing inflammation and efferocytosis rates.

From Figure 14, lower LDL levels and higher HDL levels have a positive influence on plaque development by all observed metrics, and the higher $\sigma_l$ is, the higher $\sigma_p$ has to be in order to counteract LDL's effects. High LDL influx is associated with faster plaque growth, while higher unsaturated HDL influx has the opposite effect due to its inhibition of LDL modification and monocyte recruitment, and its promotion of foam cell emigration. For high enough $\sigma_p$ and low $\sigma_l$, plaque stabilisation is achieved. High levels of HDL relative to LDL are also associated with higher proportions of live foam cells in the plaque, as RCT is able to reduce cytotoxic cholesterol loads more effectively. The increased live cell counts help to slow plaque growth, since HDL's upregulation of migratory receptors affects higher numbers of foam cells. The rates of efferocytic uptake are also influenced by the lipoprotein profile. When the lipid profile has high LDL and low HDL, lower proportions of efferocytosis-capable live foam cells lead to poor efferocytic recycling of dead material. Cases with low LDL and high HDL on the other hand have faster efferocytic uptake of dead material.

The above results suggest the categorisation of plaque lipid profiles into two regimes: a low LDL, high HDL regime with stable plaque size and healthy live foam cell counts, and an unstable high LDL, low HDL regime with mostly dead tissue and continuous plaque growth. There is also a third borderline regime between these two regions that is more sensitive to biological parameter changes.

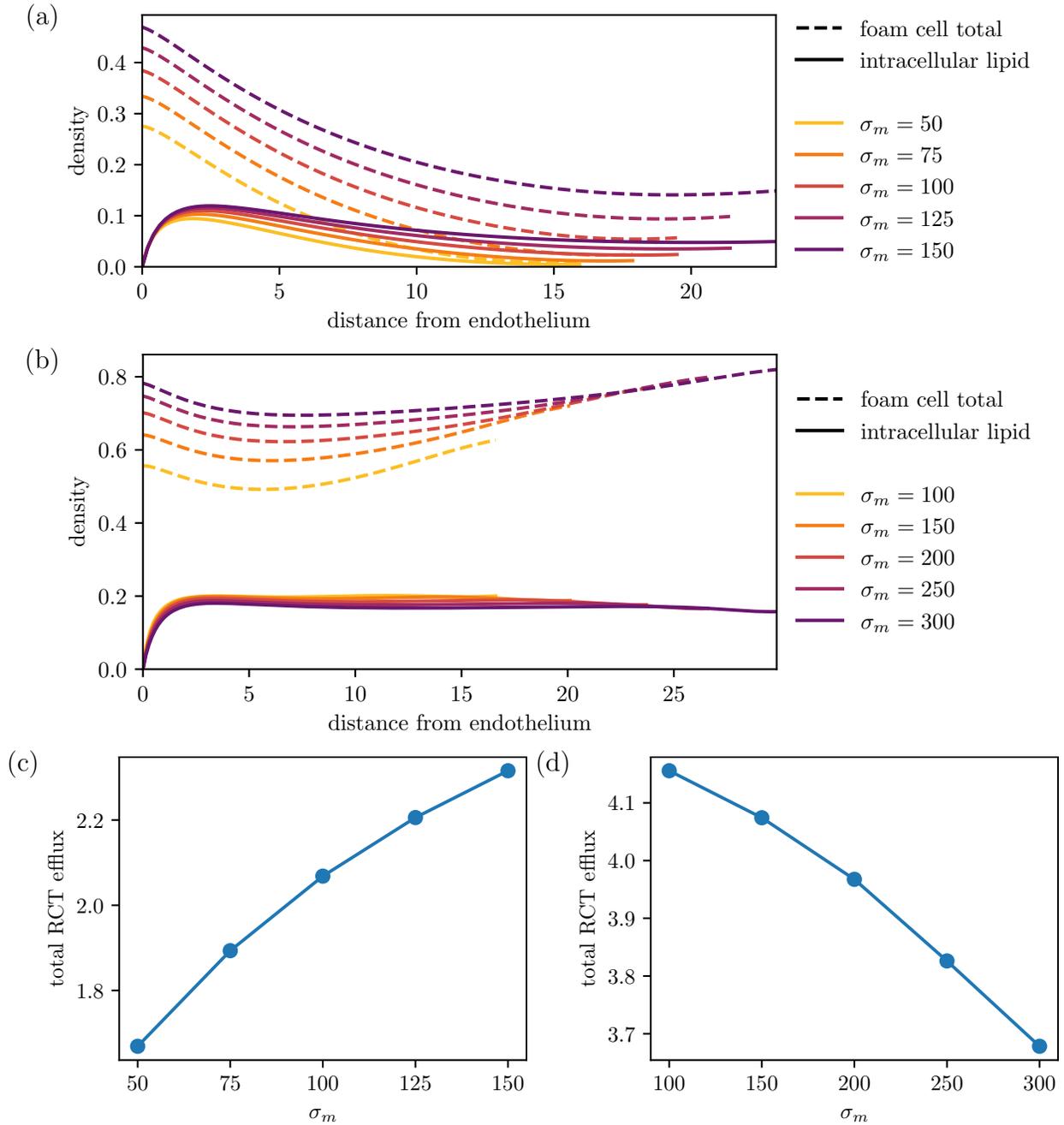

Figure 12: Comparison of (a,b) the local and intracellular cholesterol and total foam cell densities $a$ and $m + a$, and (c,d) the total cholesterol efflux rate $\int_0^R \mu_h p a \, \mathrm{d}x$, for varying inflammation rates $\sigma_m$, for plaque scenarios with (a,c) poor efferocytosis $\mu_e = 20$, or (b,d) higher efferocytosis $\mu_e = 60$. Solutions are for the simplified RCT-only HDL model (equations (38) and (39)) with fixed HDL influx and no foam cell emigration ($\sigma_p = 40$, $\sigma_e = 0$, $\sigma_l = 10$), plotted at $t = 1$.

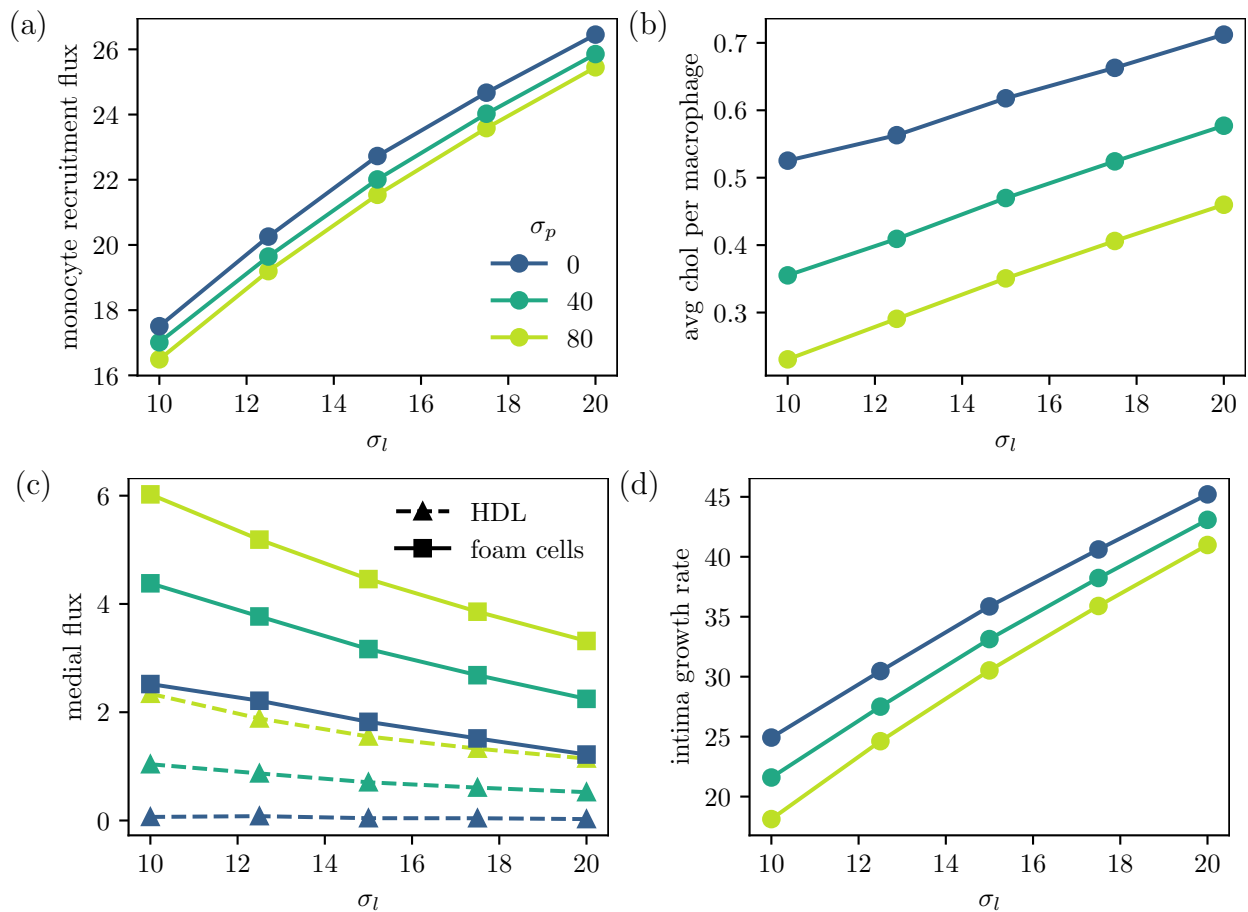

Figure 13: Comparison of various quantities for different influxes of LDL $\sigma_l$ and unsaturated HDL $\sigma_p$. Plots compare the values at $t = 1$ of (a) the endothelial monocyte recruitment flux $v_m m|_{x=0}$, (b) the average cholesterol per macrophage $\int_0^R a\, dx / \int_0^R m\, dx$, (c) the medial fluxes of HDL $v_h|_{x=R}$ and foam cells $v_m(m+a)|_{x=R}$, and (d) the intima growth rate $\frac{dR}{dt}$. Solutions are for the simplified RCT-only HDL model (equations (38) and (39)) with $\sigma_m = 100$, $\sigma_e = 10$, and $\mu_e = 40$.

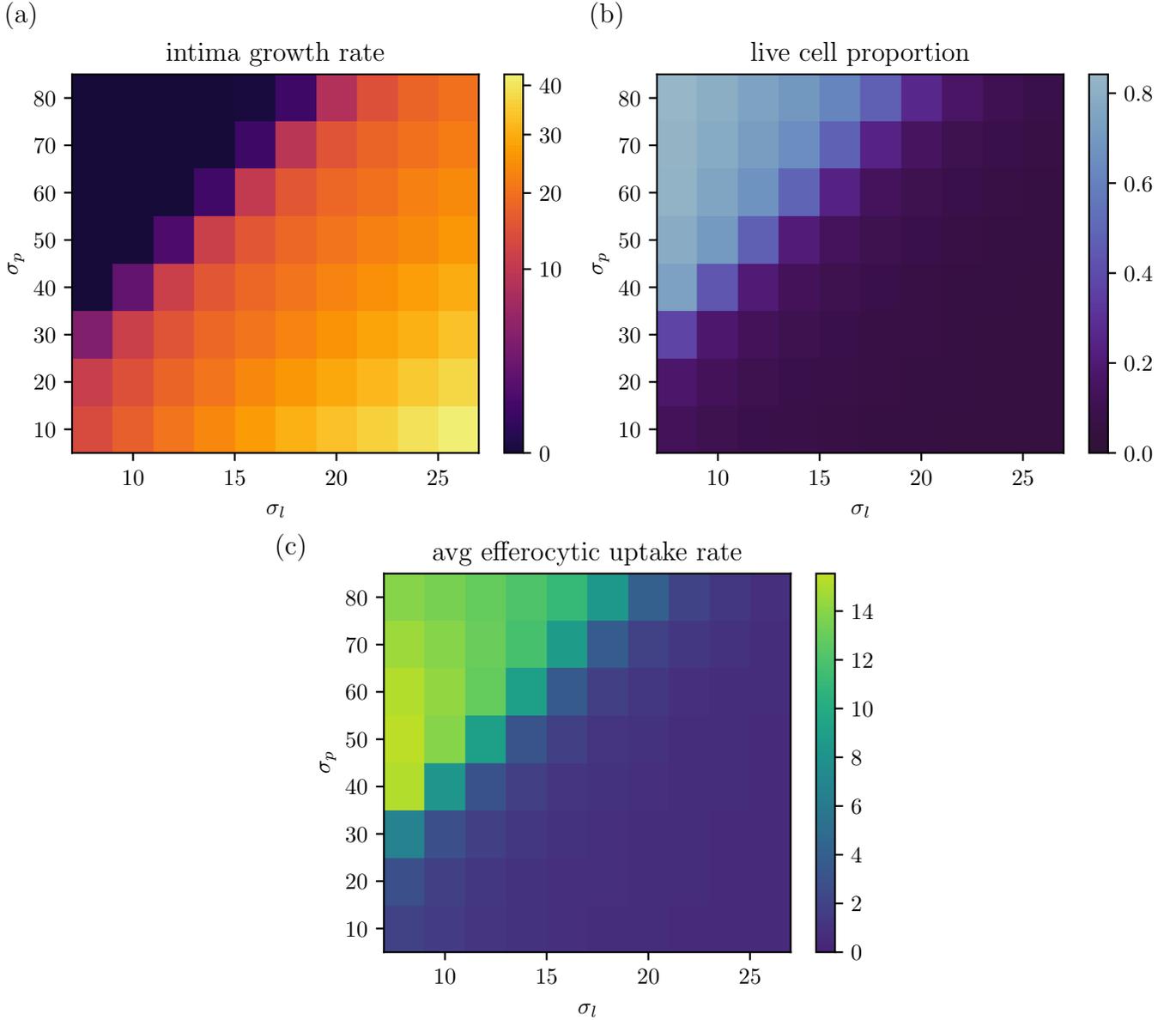

Figure 14: Heat maps of (a) the intima growth rate $\frac{dR}{dt}$, (b) the proportion of live foam cells as a fraction of total live and dead cells $\int_0^R (m+a)\,dx / \int_0^R (m+a+c+b)\,dx$, and (c) the efferocytic uptake rate averaged over all dead material $\int_0^R \mu_e m(b+c)\,dx / \int_0^R (b+c)\,dx$, for varying LDL and HDL influx rates $\sigma_l$ and $\sigma_p$. Solutions are for the full HDL model with $\sigma_m = 100$ and $\sigma_e = 30$, plotted at $t = 2$. Plot (a) uses a square root scale to emphasise values near $\frac{dR}{dt} = 0$, which represents a stable plaque size.

Figure 15(a) and (b) show how increasing the efferocytic capacity from $\mu_e = 30$ to $\mu_e = 40$ influences plaque development. For scenarios with very high LDL and low HDL where macrophages are heavily laden with cholesterol and mostly dead, increasing efferocytosis slightly does little to change things. At the other end of LDL-HDL space, stable plaques with already high live foam cell counts have less room for improvement, and increasing $\mu_e$ causes only a modest increase in live foam cell counts. For borderline cases however, increasing $\mu_e$ will boost efferocytic uptake sufficiently to prevent complete foam cell death. The boost in live cell counts has the effect of kickstarting foam cell emigration, which significantly slows plaque growth and the accumulation of dead material.

When considering the effect of increasing inflammation from $\sigma_m = 100$ to $\sigma_m = 125$ (Figure 15(c) and (d)), LDL-HDL space again separates into three distinct regimes. Here there is a dramatic difference between the unstable and borderline regimes than when efferocytosis was increased. Again, plaques situated safely in the stable low LDL, high HDL regime retain their stability when $\sigma_m$ is increased. The reduced average cholesterol loads causes a modest increase in the live foam cell proportion. For unstable plaques with high LDL and low HDL, increasing inflammation only introduces greater numbers of new macrophages which will inevitably die and remain in the plaque (as was observed in Section 3.2). In the unstable regime then, inflammation will only exacerbate plaque growth and the accumulation of dead material. In the borderline regime however, increasing the recruitment of monocytes will reduce average cytotoxic cholesterol loads enough to boost live foam cell counts significantly. This leads to enhanced emigration and slower plaque growth.

# 6 Time-varying lipid profiles and plaque regression

In this section, we use the full model from Section 2 to examine how plaques develop under time-varying LDL and HDL influxes. We use our model to simulate how decreasing LDL and increasing HDL influx parameters can induce a reduction in plaque size. We also consider how plaque growth or regression is influenced by the timing of changes in lipid profile, and how a reversion to the initial lipid profile can reverse any beneficial changes in plaque size.

## 6.1 Experimental background and model formulation

We consider four different lipid profiles, which are based qualitatively on an experimental study by Feig et al (21). In this study, aortic plaques were first grown in apolipoprotein E knockout (apoE$^{-/-}$) mice by feeding them a high fat Western diet for 16 weeks post-weaning. ApoE$^{-/-}$ mice on a high fat diet will rapidly develop plaques due to their poor ability to produce HDL, and their high levels of LDL (41). After 16 weeks, aortic plaque tissue was transplanted into recipient mice, with some tissue retained as a baseline for histological studies. The recipient mice were one of four types: apoE knockout (low HDL and high LDL), apoAI knockout (apoAI$^{-/-}$; low LDL and low HDL due to inability to synthesise apolipoprotein A-I), apoE knockout expressing the human apoAI transgene (hAI/apoE$^{-/-}$; normal HDL levels with high LDL levels), or wild type (normal HDL and low LDL). These four groups were maintained on a standard chow diet for 1 week before being euthanised for plaque analysis.

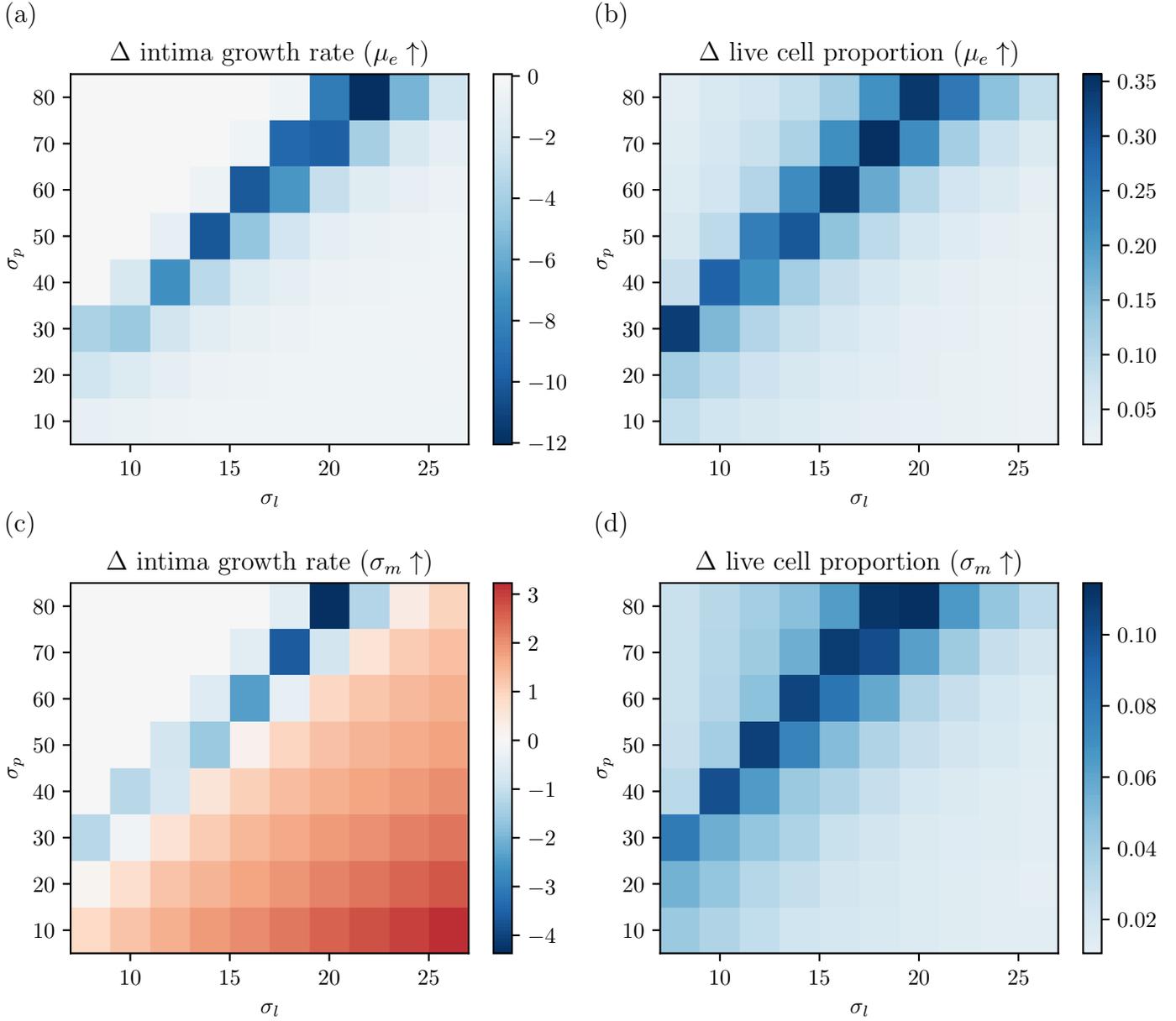

Figure 15: Heat maps of the change in the intima growth rate and live cell proportion when either the efferocytosis parameter $\mu_e$ or the monocyte recruitment parameter $\sigma_m$ are increased, for various LDL and HDL influx rates $\sigma_l$ and $\sigma_p$. In (a,b), $\mu_e$ is increased from 30 to 40, with $\sigma_m = 100$. In (c,d), $\sigma_m$ is increased from 100 to 125, with $\mu_e = 30$. Quantities plotted are (a,c) the intima growth rate $\frac{dR}{dt}$, and (b,d) the proportion of live foam cells as a fraction of total live and dead cells $\int_0^R (m + a)\, dx / \int_0^R (m + a + c + b)\, dx$, all plotted at $t = 2$. Solutions are for the full HDL model.

Compared to the apoE$^{-/-}$ group, where the plaque continued to grow post-transplant, the lower LDL apoAI$^{-/-}$ group exhibited a slight regression in plaque size. The higher HDL wild type and hAI/apoE$^{-/-}$ groups exhibited a more pronounced decrease in plaque size, particularly in the wild type group, which had both higher HDL and lower LDL levels than the baseline apoE$^{-/-}$ donor group. These results are summarised in Table 3.

Table 3: Summary of the plaque regression results from Figure 1 in Feig et al (21), where plaque-bearing aortic tissue was developed in apoE knockout mice, transplanted into one of four recipient groups, and grown for another week.

| Mouse type | Plasma LDL / HDL | | Change in plaque size vs baseline |
| --- | --- | --- | --- |
| apoE$^{-/-}$ | high / low | (control) | increase |
| apoAI$^{-/-}$ | low / low | (decrease / no change) | small decrease |
| hAI/apoE$^{-/-}$ | high / normal | (no change / increase) | moderate decrease |
| wild type | low / normal | (decrease / increase) | large decrease |

We simulate the effects of changing lipid profiles using the full six-phase model from Section 2. We consider four different sets of LDL and HDL influx values $\sigma_l, \sigma_p$ that represent the four mouse groups, which are given in Table 4. We model three different scenarios:

1. A persistent change from an apoE$^{-/-}$ lipid profile to one of the four lipid profiles at a fixed time (where the apoE$^{-/-}$ to apoE$^{-/-}$ group effectively undergoes no change).

2. A persistent change from an apoE$^{-/-}$ to a wild type lipid profile at different times.

3. A temporary change from an apoE$^{-/-}$ to a wild type lipid profile at different times, reverting to an apoE$^{-/-}$ profile after a fixed "therapy" duration.

Lipoprotein influxes $\sigma_l$ and $\sigma_p$ are transitioned smoothly using a sigmoid function:

$$\sigma(t) = \sigma_1 + (\sigma_2 - \sigma_1)\frac{1}{1 + e^{-(t-t_1)/0.05}} + (\sigma_3 - \sigma_2)\frac{1}{1 + e^{-(t-t_2)/0.05}}, \quad (40)$$

denoting successive changes from a base value of $\sigma_1$ to $\sigma_2$ and then $\sigma_2$ to $\sigma_3$, centred at times $t_1$ and $t_2$ respectively ($\sigma_2 = \sigma_3$ for scenarios 1 and 2). Remaining parameters are given in Table 2, with moderate inflammation ($\sigma_m = 150$) and poor efferocytosis ($\mu_e = 30$).

## 6.2 Results

Changes in lipid profile have an immediate effect on plaque growth. Figure 16(a) compares the total endothelial influx of monocytes and modLDL for scenario 1, where each case is simultaneously transitioned from an apoE$^{-/-}$ type lipid profile (high LDL, low HDL) to one of the four profiles in Table 4. Here, the lower LDL apoAI$^{-/-}$ case sees a decrease in the rate at which new material enters, due to both the decrease in LDL and the resulting decrease in the inflammatory monocyte response. The higher HDL hAI/apoE$^{-/-}$ case also sees a reduction in the total endothelial influx, due to HDL directly inhibiting the entry of modLDL and monocytes. The wild type group sees the most dramatic drop due to the combined effects of both lower LDL and higher HDL. Figure 16(b) compares the total material efflux through the medial boundary for each case. Both high HDL groups see a significant increase in material removal due to the increased amounts of HDL that remove cholesterol from the plaque, and the increased upregulation of foam cell emigration. From Figure 16(c), the net effect is that the cases that undergo changes to a lower LDL or higher HDL lipid profile all exhibit a noticeable slowing in plaque growth when compared to the apoE$^{-/-}$ baseline case. The most significant change is observed in the wild type profile, which has both lower LDL and higher HDL.

Table 4: Lipid influx parameter values used to represent the four mouse groups in Table 3.

| Mouse type | $\sigma_l$ | $\sigma_p$ |
|---|---|---|
| apoE$^{-/-}$ | 25 | 10 |
| apoAI$^{-/-}$ | 10 | 10 |
| hAI/apoE$^{-/-}$ | 25 | 100 |
| wild type | 10 | 100 |

The efficacy of changes to the lipid profile depends on the timing of the change. Figure 17 tracks the evolution of plaques in scenario 2, where each case is switched from an apoE$^{-/-}$ type lipid profile to a wild type profile at different times ($t_1 = 0.3$, 0.4, and 0.5). All the cases see an immediate slowing in plaque growth. In the $t_1 = 0.3$ case, the plaque begins to regress in size almost immediately. The $t_1 = 0.4$ case also regresses in size, but with some delay compared to the $t_1 = 0.3$ case. For the $t_1 = 0.5$ case, plaque growth quickly slows, but does not exhibit regression within the observed timeframe. This delayed action is apparent when examining plaque behaviour near the media. From Figure 17(b), changing to a wild type lipid profile results in an immediate decrease in the influx of new modLDL and macrophages through the endothelium. This change is identical for all three cases. Foam cell emigration and HDL clearance at the medial boundary also increase after the change. Unlike at the endothelium, however, the increase in medial effluxes happens more gradually, and is less pronounced for the $t_1 = 0.5$ case than the cases with earlier intervention.

The importance of timing on the effectiveness of modifying lipid profiles can be attributed to differences in the amount of plaque material that has already accumulated. It was observed in Figure 8(d) and Figure 9(a) that HDL's cholesterol efflux action is mostly localised near the endothelium. The benefits of the increased HDL influx therefore affect newer foam cells more than older foam cells deeper inside the plaque. This existing backlog of foam cells will initially emigrate slowly due to them developing under conditions of higher LDL and lower HDL (meaning fewer live cells and greater inhibition of emigration due to higher cholesterol loads, and lower upregulation of emigration by HDL). The composition of plaque tissue near the medial boundary will only start to reflect the new lipid profile as this existing material is cleared. Hence increased medial efflux will take longer to begin when the change

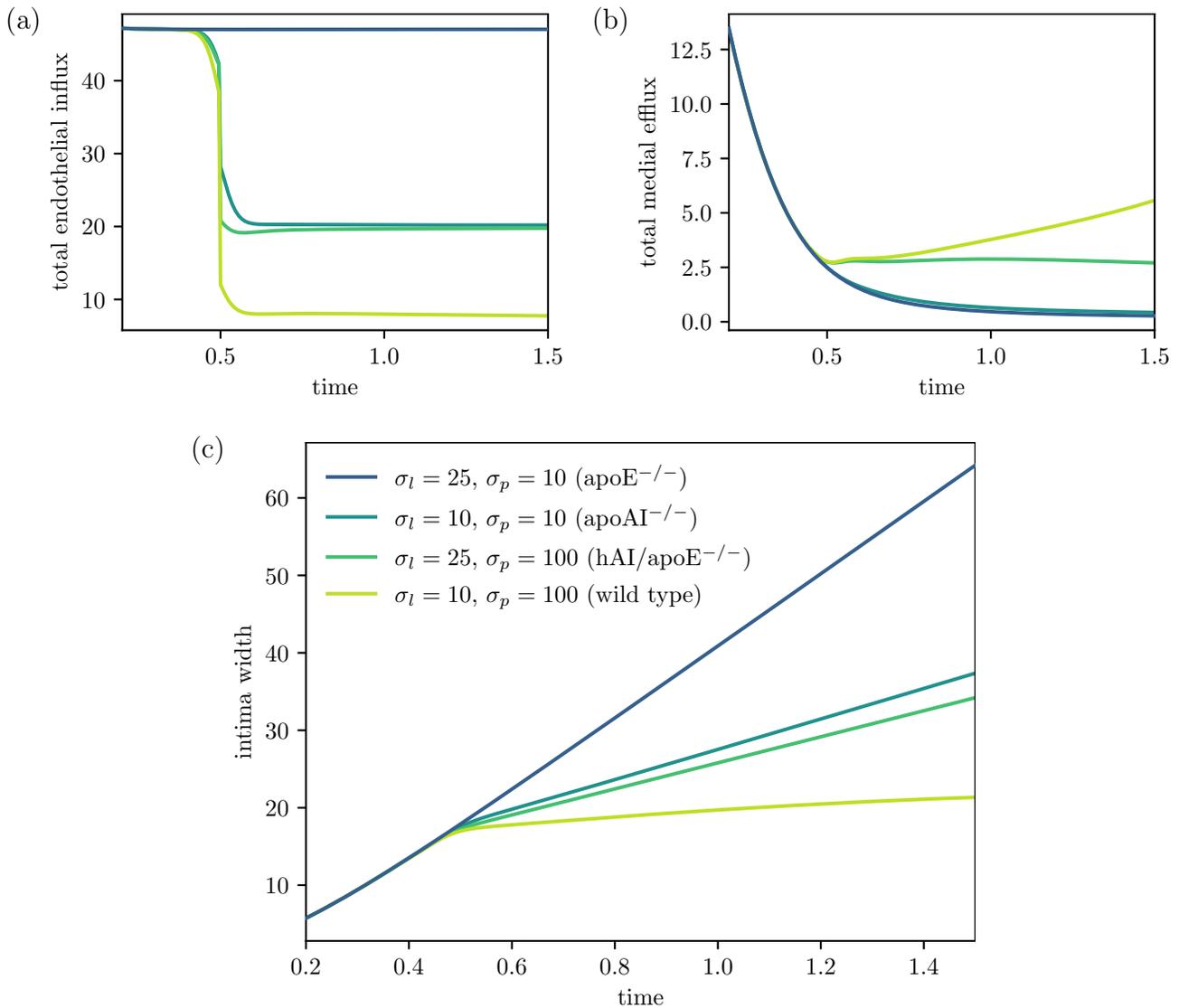

Figure 16: Comparisons of the time evolution of (a) the total material influx at the endothelium $(v_m m + v_l l)|_{x=0}$, (b) the total material efflux at the media $(v_m(m+a) + v_h h)|_{x=R}$, and (c) the width of the intima $R(t)$ for simulation 1, where an apoE$^{-/-}$ lipid profile is either maintained, or switched at $t_1 = 0.5$ to one of three other lipid profiles (Table 4).

is administered later.

Temporary changes in lipid profile are also unlikely to be effective at improving plaque health in the long term. Figure 18 looks at scenario 3, where two plaques were switched from an apoE$^{-/-}$ to a wild type lipid profile at different times, and reverted to an apoE$^{-/-}$ profile after an equal duration of "therapy" time. Both cases exhibit reduced modLDL and monocyte influxes and increased foam cell and HDL clearance following the first change. The early intervention case also exhibits regression in plaque size, unlike the later intervention case. After the second change, the total modLDL and monocyte influxes quickly revert to pre-treatment rates for both cases. This causes a reversion to the original higher rate of plaque growth. While the total medial HDL and foam cell effluxes remain higher than pre-treatment levels immediately following the reversion, they eventually fall again. As in scenario 2 and Figure 17 however, the early intervention case still sees a greater reduction in plaque growth during the period when the therapy is in effect.

# 7 Discussion

In this paper, we presented a free boundary multiphase model for early atherosclerotic plaque development under the influence of both LDL and HDL. A key feature of this model is the explicit inclusion of continuous spatial heterogeneity in intracellular cholesterol levels.

In the HDL-free model, increasing macrophage efferocytic capacity has a uniformly beneficial effect on plaque development. Increased efferocytosis causes lower dead cell counts and slower plaque growth, due to the greater availability of live foam cells that are able to emigrate with the cholesterol they carry. Although the model is a spatial continuum model that does not measure individual cell lifetimes, the increased rates of efferocytic recycling suggest improved cell turnover, where apoptotic cells remain in the plaque for less time before being efferocytosed, thus reducing the chances of cell necrosis. This is consistent with observations in non-spatial plaque growth models (24; 13; 38) and in experimental studies (60; 33), where defective efferocytosis is linked with aggravated plaque growth and increased necrotic material accumulation.

Inflammation has a more interesting interplay with cholesterol-induced death. Increasing inflammation can exacerbate plaque growth by introducing more new cells to the plaque, some of which will contribute to total dead cell counts. However, the increase in monocyte recruitment also has the effect of diluting total cholesterol loads over a larger quantity of macrophages. In a more moderate plaque scenario, this reduces cholesterol-induced cell death and boosts the ratio of live to dead foam cells, which in turn boosts emigration rates and improves cell turnover. In a pathological scenario with highly defective efferocytosis however, increasing inflammation only introduces more macrophages that inevitably die. Advanced plaques tend to fall into the latter category (33), which justifies the therapeutic focus on anti-inflammatory drugs over pro-inflammatory treatments for atherosclerosis (15).

In the model with HDL, we find that reverse cholesterol transport is highly beneficial in improving plaque health. By reducing foam cell cholesterol loads, HDL inhibits cholesterol-induced cell death, thereby improving efferocytic cell uptake, and boosting live cell counts and emigration rates. These improvements are observed both in the simplified RCT-only HDL model from Section 4 that ignores HDL's upregulation of migratory receptors, and in the full model considered in Section 5. This is consistent with studies involving mice with defective expression of the ABCA1 transporter (63), which is a key transporter involved in cholesterol efflux to HDL. These mice exhibit larger plaques with higher foam cell cholesterol content and impaired macrophage emigration. It is unclear whether the reduced migration is due to increased cholesterol loading or other factors, since other studies in mice suggest that ABCA1 deletion also inhibits macrophage emigration via other pathways (46).

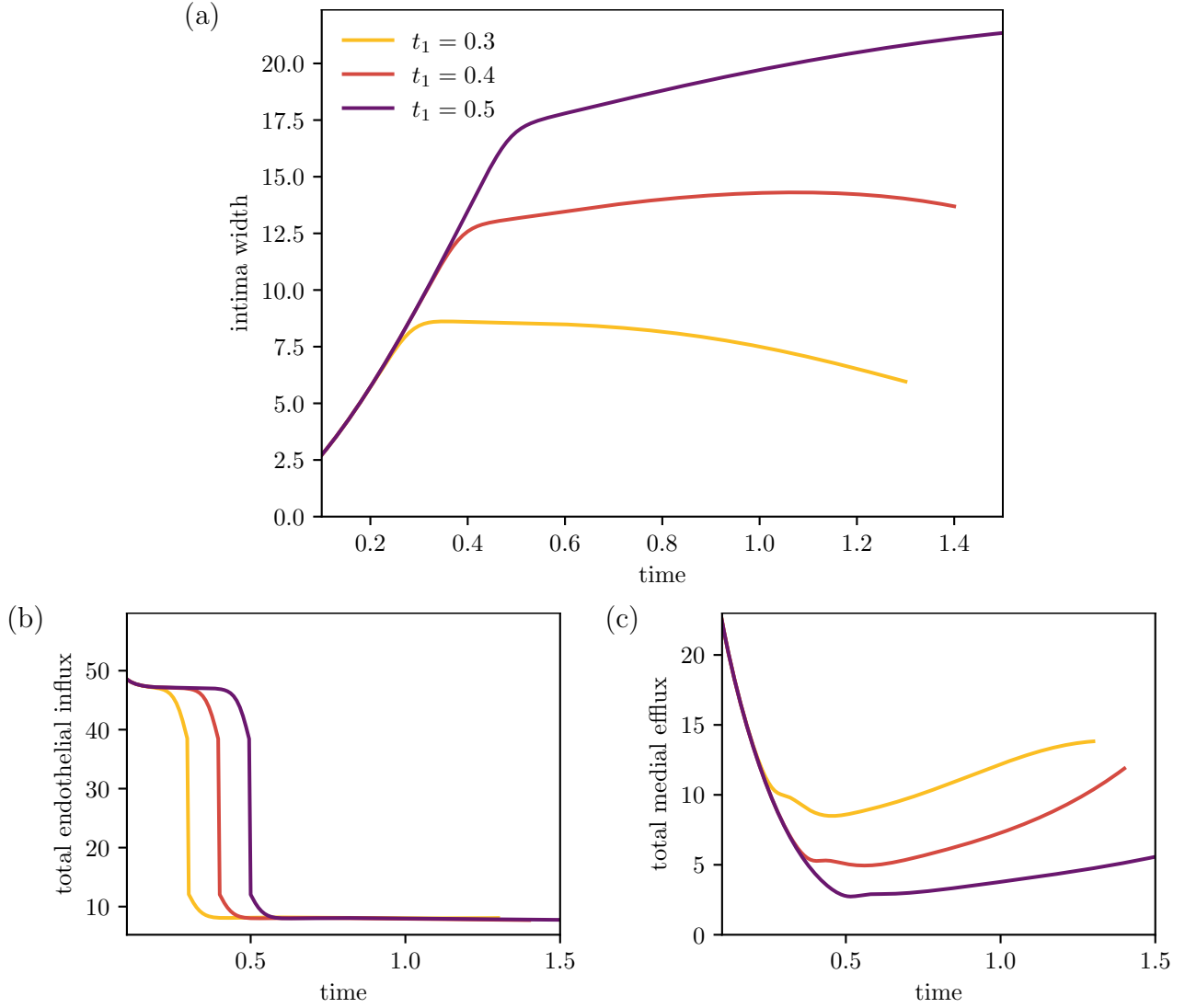

Figure 17: Comparisons of the time evolution of various quantities for plaques switched from an apoE$^{-/-}$ type ($\sigma_l = 25$, $\sigma_p = 10$) to a wild type ($\sigma_l = 10$, $\sigma_p = 100$) lipid profile at different times ($t_1 = 0.3$, $0.4$, and $0.5$). Quantities plotted are (a) the width of the intima $R(t)$, (b) the total material influx at the endothelium $(v_m m + v_l l)|_{x=0}$, and (c) the total material efflux at the media $(v_m(m + a) + v_h h)|_{x=R}$.

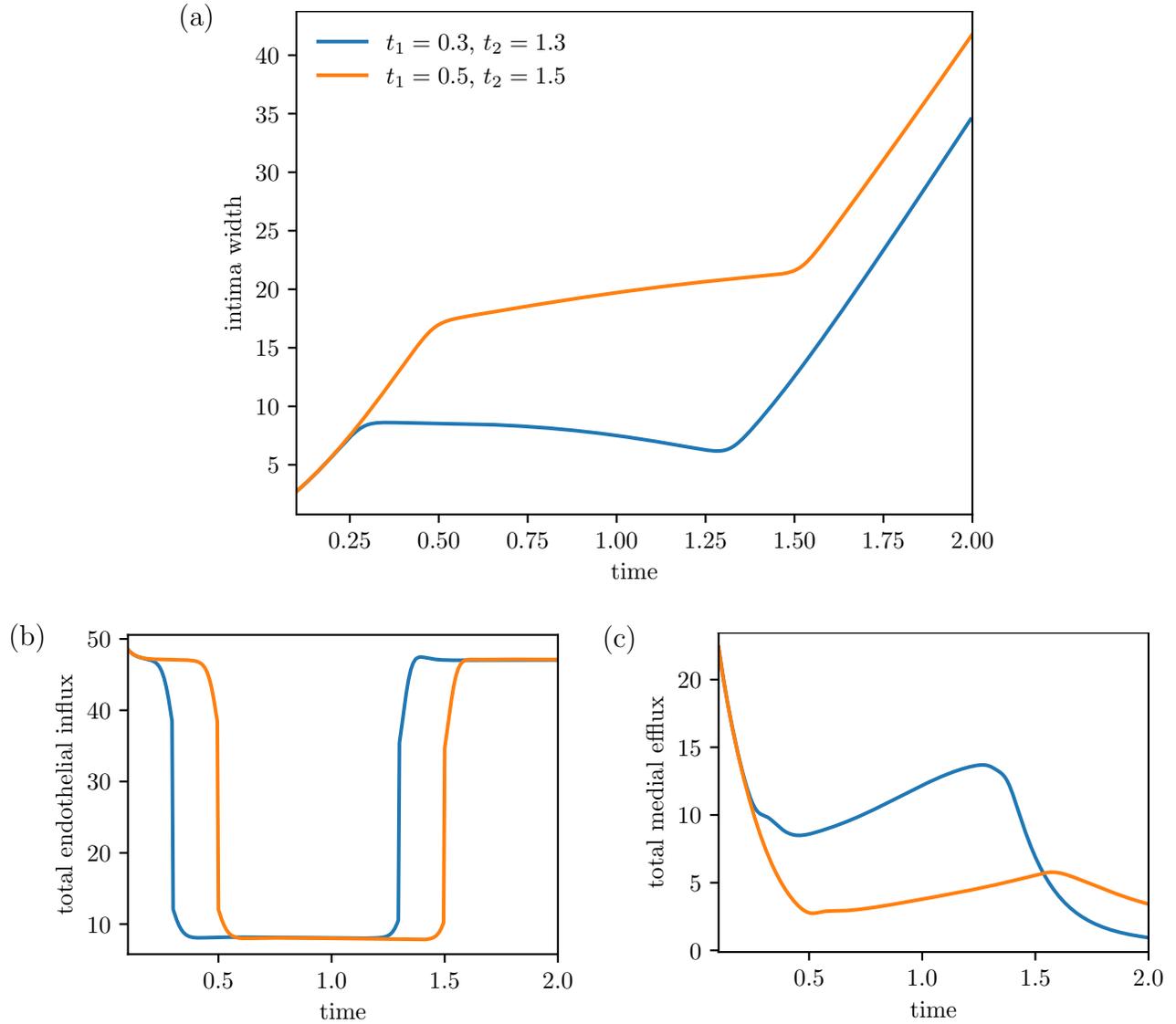

Figure 18: Comparisons of the time evolution of various quantities for plaques switched from an apoE$^{-/-}$ type ($\sigma_l = 25$, $\sigma_p = 10$) to a wild type ($\sigma_l = 10$, $\sigma_p = 100$) lipid profile at different times ($t_1 = 0.3$ and $0.5$), and then reverted to the apoE$^{-/-}$ type profile at time $t_2 = t_1 + 1$. Quantities plotted are (a) the width of the intima $R(t)$, (b) the total material influx at the endothelium $(v_m m + v_l l)|_{x=0}$, and (c) the total material efflux at the media $(v_m(m + a) + v_h h)|_{x=R}$.

Plaque development seems to fall under one of two broad regimes depending on the blood lipid profile. High LDL and low HDL influxes put plaques in an unstable regime characterised by rapid plaque growth, low proportions of live foam cells, and high quantities of apoptotic or necrotic material with negligible efferocytic recycling. Plaques with low LDL and high HDL fall under a stable regime, where plaques reach steady state and remain small, with higher proportions of live foam cells relative to dead material, and healthier rates of efferocytic uptake. This agrees with existing models for HDL in plaques (19; 27; 59), where high LDL and low HDL is associated with sustained growth, and low LDL and high HDL with steady state plaques. Our model is also broadly consistent with experimental studies, where high LDL and low HDL have long been understood to have an atherogenic effect (2; 62; 21). Interestingly, our model contains a borderline regime between the unstable and stable regions where the plaque is more sensitive to changes in efferocytic activity or inflammation. In particular, increasing inflammation in the unstable regime will only exacerbate plaque growth by supplying new cells that inevitably join the necrotic core. In the borderline regime, however, the dilution of average cholesterol loads will reduce the death rate enough to tip plaques over into the stable regime.

Our model results underscore the importance of early intervention when timing anti-atherogenic therapies. This is consistent with studies in mice where anti-atherogenic treatments are administered at different stages of plaque progression to multiple groups of otherwise similar mice (17; 10). One study by Chyu et al (17) involves immunising apoE$^{-/-}$ mice with LDL to bolster the recognition of LDL by T-cells. Immunisation was carried out at either 7 or 20 weeks of age, with both groups maintained on the same diet and sacrificed at the same time. Mice in the late group developed larger plaques and had lower levels of oxidised LDL antibodies. The diminished effectiveness of therapies when applied to more advanced plaques may also explain why human drug trials tend to be less conclusive than mouse model studies. Mice used in studies bear a very specific genetic profile and have their diets controlled carefully from birth. Plaques grown in mice therefore tend to be at very similar stages of progression, and will respond similarly to treatment. In contrast, human test subject groups have a wider spread of genetic profiles and life histories that lead to varying degrees of plaque progression, and hence a wider spread of outcomes.

Our results also suggest that anti-atherogenic therapies must be sustained in order to be effective, or must induce some sort of persistent change. The reverse has been observed in animal studies (16), where temporarily feeding animals a high cholesterol diet and then removing them from the diet again results in plaque regression. Relevant data in human trials is scarce however, due to a lack of follow-up studies that check up on subjects again some time after treatment has been discontinued.

This model is the first spatial PDE model for atherosclerosis that includes continuous heterogeneity in foam cell lipid content. Lipid-structured models such as those by Ford et al (25) and (14) explicitly include intracellular lipid content as an independent variable. In these models, the rates of processes such as cell death, emigration, and phagocytosis have a functional dependence on the intracellular lipid variable. Since these models lack spatial structure, however, discussions regarding plaque growth or regression must use total cell counts or similar quantities as a proxy for plaque size. A number of spatially structured plaque models incorporate free boundaries, including multiphase models (1; 27) and continuum mechanical models (58; 23). The free boundary allows them to directly quantify plaque growth via the spatial domain size. These models treat new macrophages and cholesterol-engorged foam cells as separate but phenotypically homogeneous species, however. By separating macrophages from their intracellular cholesterol content, our six-phase model is able to incorporate some of the strengths of lipid-structured PDE models into a spatial PDE framework. In particular, our model framework can capture continuous phenotypic changes induced by cholesterol loading without requiring another cell species. At the same time, having spatial structure means that movement processes and local interactions (such as emigration and efferocytosis respectively) can be modelled in a way that respects local mass conservation and spatial crowding.

Our model treats the endogenous lipid content of macrophages as part of the macrophage phase $m(x,t)$. Real cells contain some amount of endogenous cholesterol, which combines with the cholesterol from ingested modLDL. Our model could be extended to include endogenous cholesterol, via a nonzero endothelial flux term for the intracellular cholesterol phase $a(x,t)$. If death rates are low for macrophage foam cells with typical endogenous cholesterol levels, we would expect the model to produce largely similar results to those presented here. Results from lipid-structured models however (24; 14) suggest that the endogenous lipid content of incoming monocytes can act as a significant source of cholesterol in addition to LDL. Furthermore, cannibalistic macrophage efferocytosis can lead to the accumulation of highly cytotoxic cholesterol levels within individual macrophages. An open question is whether these effects can be captured in a spatial model.

One shortcoming of the continuous multiphase modelling approach used by our model is that it does not directly account for cell numbers. This has implications for efferocytosis, since our model does not distinguish between a group of smaller macrophages with smaller individual cholesterol loads, and a single large macrophage with a larger cholesterol load but the same intracellular cholesterol volume proportion. We may however assume that macrophages are also undergoing proliferation into smaller macrophages, which has been observed in plaques (51) and is believed to be important, but is difficult to model in a spatially averaged mass conservative framework. Proliferation would limit overall cell sizes and prevent excessive cholesterol biomagnification, and this has been demonstrated in lipid-structured models (13). Modelling the effects of proliferation and cholesterol biomagnification in a spatially structured model requires a different approach, such as hybrid spatial-lipid structured models, or agent-based spatial models that allow variable levels of intracellular cholesterol.

# 8 Conclusions

In this paper, we present a novel multiphase model for early plaque formation that allows continuous variation in intracellular cholesterol content. This is the first spatial model with continuous phenotypic changes in plaque macrophage behaviour, where these changes are driven by internalised cholesterol. The model directly address questions about how plaque growth and regression is affected by the spatial distribution of intracellular cholesterol. We used this model to explore how cholesterol accumulation, inflammation, and efferocytosis influence plaque structure and growth. We used the model to show how lipid dynamics, driven by LDL and HDL influence plaque development, both for static LDL-HDL blood lipid profiles and for time-varying lipid profiles. The model forms a good foundation for more complex dynamical models for early plaque development that account for more varied cell behaviours and types.

# Appendix A  Nondimensional model

The full nondimensional model is obtained by applying the nondimensionalisation in Section 2.1 to the model in Section 2. The nondimensional model consists of the continuity equations

$$\frac{\partial u}{\partial t} = -\frac{\partial}{\partial x}(J_u + vu) + s_u = -\frac{\partial}{\partial x}(v_u u) + s_u \,, \tag{41}$$

for species $u = m, c, l, a, b, h, p$, with a no-voids condition

$$m + c + l + a + b + h = 1 \,, \tag{42}$$

mass exchange terms

$$\begin{aligned}
s_m &= \phantom{-}+\mu_e mc - \mu_a \bar{a} m && , & (43)\\
s_c &= \phantom{+}-\mu_e mc + \mu_a \bar{a} m && , & (44)\\
s_l &= -\mu_p ml && , & (45)\\
s_a &= +\mu_p ml + \mu_e mb - \mu_a \bar{a} a - \mu_h ap && , & (46)\\
s_b &= \phantom{+}-\mu_e mb + \mu_a \bar{a} a && , & (47)\\
s_h &= \phantom{+}+\mu_h ap && , & (48)\\
s_p &= \phantom{+}-\mu_h ap - \delta_p p && , & (49)\\
&&&& (50)
\end{aligned}$$

flux terms

$$J_m = -D_m \frac{\partial m}{\partial x} + \chi_l m \frac{\partial l}{\partial x} + \chi_c m \frac{\partial}{\partial x}(c+b) \,, \tag{51}$$

$$v_a = v_m \,, \tag{52}$$

$$J_u = -D_u \frac{\partial u}{\partial x} \quad \text{for } u = l, c, b, h, p \,, \tag{53}$$

boundary conditions at $x = 0$

$$v_m m \Big|_{x=0} = \frac{\sigma_m l}{1 + \sigma_p/\sigma_{p,\text{inh}}} \Big|_{x=0}, \tag{54}$$

$$v_l l \Big|_{x=0} = \frac{\sigma_l}{1 + \sigma_p/\sigma_{p,\text{inh}}}, \tag{55}$$

$$v_u u \Big|_{x=0} = 0 \quad \text{for } u = a, c, b, h, \tag{56}$$

boundary conditions at $x = R(t)$

$$(v_m - R')m \Big|_{x=R} = \frac{p+h}{(p+h) + h_{\text{egr}}} \sigma_e (1 - \bar{a}) m \Big|_{x=R}, \tag{57}$$

$$(v_h - R')h \Big|_{x=R} = \sigma_h h, \tag{58}$$

$$(v_p - R')p \Big|_{x=R} = \sigma_h p, \tag{59}$$

$$(v_u - R')u \Big|_{x=R} = 0 \quad \text{for } u = l, c, b, \tag{60}$$

and initial conditions

$$m(x, 0) = 1, \tag{61}$$

$$u(x, 0) = 0 \quad \text{for } u = a, l, c, b, h, p, \tag{62}$$

$$R(0) = 1, \tag{63}$$

where the mixture velocity is

$$v = \left(1 + \frac{a}{m}\right)\left(D_m \frac{\partial m}{\partial x} - \chi_l m \frac{\partial l}{\partial x} - \chi_c m \frac{\partial}{\partial x}(c+b)\right) \\ + D_l \frac{\partial l}{\partial x} + D_c \frac{\partial}{\partial x}(c+b) + D_h \frac{\partial h}{\partial x} + \frac{\sigma_m l + \sigma_l}{1 + \sigma_p/\sigma_{p,\text{inh}}} \Big|_{x=0}, \tag{64}$$

and the domain grows with rate

$$\frac{dR}{dt} = \frac{\sigma_m l + \sigma_l}{1 + \sigma_p/\sigma_{p,\text{inh}}} \Big|_{x=0} - \frac{p+h}{(p+h) + h_{\text{egr}}} \sigma_e (1 - \bar{a})(m + a) \Big|_{x=R} - \sigma_h h \Big|_{x=R}. \tag{65}$$